\documentclass[10pt,journal,compsoc]{IEEEtran}

\IEEEoverridecommandlockouts

\usepackage{graphics}
\usepackage{epsfig}
\usepackage{multicol}
\usepackage{multirow}
\usepackage{amsmath}
\usepackage{amssymb}
\usepackage{colortbl}
\usepackage{siunitx}
\usepackage[version=4]{mhchem}
\usepackage{todonotes}
\usepackage{dirtytalk}
\usepackage[normalem]{ulem}

\usepackage{color}

\newcommand{\review}[1]{{\color{black}{#1}}}
\newcommand{\secondreview}[1]{{\color{black}{#1}}}
\newcommand{\finalreview}[1]{{\color{black}{#1}}}
\graphicspath{{./Figures/}}

\begin{document}
%%%%%%Added formatting for ArXiv submission
\onecolumn
\textcopyright~2019 IEEE.  Personal use of this material is permitted.  Permission from IEEE must be obtained for all other uses, in any current or future media, including reprinting/republishing this material for advertising or promotional purposes, creating new collective works, for resale or redistribution to servers or lists, or reuse of any copyrighted component of this work in other works.
\newpage
\twocolumn
\title{Understanding Continuous \review{and Pleasant} Linear Sensations \finalreview{on the Forearm} from a Sequential Discrete Lateral Skin-Slip Haptic Device}
%\title{Understanding Continuous \review{and Pleasant} Linear Sensations from a Sequential Discrete Lateral\\ Skin-Slip Haptic Device}
\author{Cara M. Nunez,~\IEEEmembership{Student Member,~IEEE,} Sophia R. Williams,~\IEEEmembership{Student Member,~IEEE,}\\ Allison M. Okamura,~\IEEEmembership{Fellow,~IEEE,} and Heather Culbertson,~\IEEEmembership{Member,~IEEE}% <-this % stops a space
\thanks{C.\ M.\ Nunez, S.\ R.\ Williams, and A.\ M.\ Okamura are with the Departments of Bioengineering, Electrical Engineering, and Mechanical Engineering at Stanford University, Stanford, CA 94305.  (email: \{nunezc; sophiarw; aokamura\}@stanford.edu).} 
\thanks{H.\ Culbertson is with the Department of Computer Science at University of Southern California, Los Angeles, CA 90089 (email: hculbert@usc.edu).}
\thanks{This work was supported in part by a grant from Facebook, Inc., the National Science Foundation Graduate Fellowship Program, and National Science Foundation grant 1830163.}}

\markboth{IEEE Transactions on Haptics,~Vol.~XX, No.~XX, August~2019}%
{Nunez \MakeLowercase{\textit{et al.}}: Understanding Continuous Linear Sensations from a Sequential Discrete Lateral Skin-Slip Haptic Device}

\IEEEtitleabstractindextext{
\begin{abstract}
A continuous stroking sensation on the skin can convey messages or emotion cues. We seek to induce this sensation using a combination of illusory motion and lateral stroking via a haptic device. Our system provides discrete lateral skin-slip on the forearm with %five
rotating tactors, which independently provide lateral skin-slip in a timed sequence. We vary the sensation by changing the \secondreview{angular velocity} % duration of tactor rotation
and delay between adjacent tactors, such that the apparent \review{speed} %velocity
of the perceived stroke ranges from 2.5 to 48.2~cm/s. We investigated which actuation parameters create the most pleasant and continuous sensations through a user study with 16 participants. On average, the sensations were rated by participants as both continuous and pleasant. The most continuous and pleasant sensations were created by apparent \review{speeds} %velocities
of 7.7 and 5.1~cm/s, respectively. We also investigated the effect of spacing between contact points on the pleasantness and continuity of the stroking sensation, and found that the users experience a pleasant and continuous linear sensation even when the space between contact points is relatively large (40~mm). Understanding how sequential discrete lateral skin-slip creates continuous linear sensations can influence the design and control of future wearable haptic devices.
\end{abstract}
}

\maketitle
%\thispagestyle{empty}
%\pagestyle{empty}

%\enlargethispage{0.2in}
\IEEEraisesectionheading{\section{Introduction}
\label{intro}}
%\hmc{Paragraph motivating why we are trying to create a stroking sensation. Where is this sensation encountered in daily life, and how can it be used practically in haptic devices?}
\IEEEPARstart{T}he sensation of stroking along the arm is a natural, pleasant sensation common in social touch that can also be exploited for conveying information such as simple messages\review{, interactions in virtual reality environments,} or directional cues. While human social touch is complex, a simple stroking sensation has been shown to be able to convey multiple emotions, including comfort, love, and sadness~\cite{hertenstein2009communication}. Haptic social touch aims to mimic the gestures that humans use in social touch interactions through electromechanical devices.
%human-human touch, which involves several different types of stroking patterns such as those to relay comfort, love, or sadness~\cite{jones1985naturalistic,hertenstein2009communication}. 
Although touch is the primary nonverbal means of communication of emotion between humans~\cite{app2011nonverbal}, haptic technology currently lacks similarly meaningful social touch signals~\cite{vanerp2015social}. Gaining a stronger understanding of how we can create more realistic haptic displays that imitate human touch will inform wearable haptic device design and improve virtual communication between humans separated by a distance~\cite{haans2006mediated} and between a human and robot~\cite{huisman2017social}. 

%\hmc{Paragraph discussing design considerations of wearable devices? Power, weight, etc. from the Pacchierotti paper.}
A major advantage of wearable haptic devices is that their reduced form factor enables the possibility to receive haptic feedback in a variety of locations or while moving, as opposed to the user needing to remain tethered to a set location as with world-grounded devices. However, the design of wearable haptic devices poses many challenges. Specifically, designers must consider form factor, weight, impairment, and comfort when determining how they will design and actuate their device. An optimal wearable haptic device would be small and compact, lightweight, comfortable, and naturally fit the human body without impairing it or interfering with normal actions and functions~\cite{Pacchierotti2017wearablereview}. At the heart of these design considerations is the choice of actuators. The actuators are usually the bulkiest and heaviest components in a haptic device and thus can greatly affect its success in serving as a wearable haptic device. Designers must choose actuators that fit the electromechanical parameters (force, power, precision and resolution, bandwidth, workspace, and degrees of freedom) which can create the desired sensation, but also keep in mind the design considerations of form factor, weight, impairment, and comfort. Instead of simply determining the optimal trade-offs, designers can use techniques like haptic illusions to use small, lightweight actuators to create sensations which typically require more mechanically robust actuators. Here we aim to determine specifications for creating the illusion of continuous lateral motion using a series of discrete lateral skin-slips. Although the device presented in this paper is not wearable, the results gathered from our studies can be applied to minimizing the form-factor of a lateral sensation device to satisfy the constraints discussed above.

%\hmc{Paragraph discussing why it is difficult to build a device to create a long stroking sensation. Discussion of why we are pursuing haptic illusions.}

Toward the goal of displaying a continuous stroking sensation, a variety of haptic displays have previously been investigated. Most commonly, continuous lateral motion has been directly applied to the skin to create this type of sensation~\cite{eichhorn2008stroking,knoop2015tickler}. However, stroke length is limited in these direct stimulation devices. It is difficult to create long stroking sensations with a wearable haptic device using lateral motion because it requires complex actuation and mechanical design, and in turn is often heavy and bulky. As an alternative, the use of haptic illusions allows designers to use less complicated actuation techniques, making them more appropriate for use in wearable devices. Researchers have begun investigating the use of vibration sequences to create the illusion of lateral motion along the skin~\cite{israr2011tactile,huisman2016simulating}. Recently, we showed that normal indentation sequences can also be used to successfully create the illusion of lateral motion along the skin~\cite{VCstudy}. As a follow-up to this previous work, here we use discrete lateral skin-slip because it combines the benefits of direct lateral motion and illusory lateral motion. \review{Our device is meant to convey social touch cues in which stroking motions are used, such as comfort and affection, and therefore we focus on investigating both the continuity and pleasantness of the sensation.}

%\hmc{Paragraph that lays out what we are presenting in this paper. "This paper presents... In Section 3 we... In Section 4 we..."}

\begin{figure}[!t]
	\centering
	\includegraphics[width=\columnwidth]{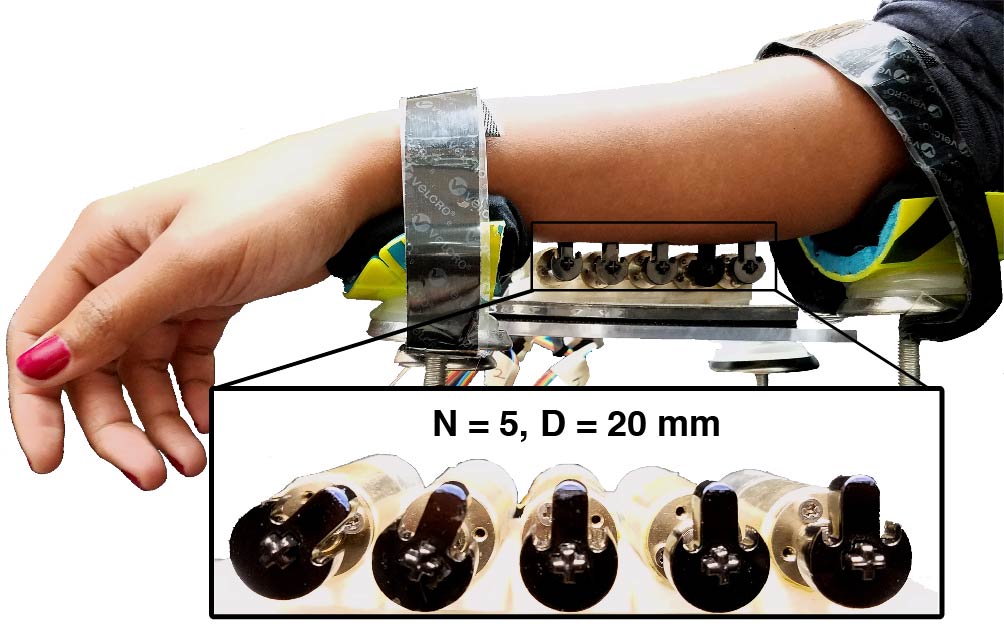}
	\vspace{-0.25in}
	\caption{Device for creating the sensation of continuous motion on the forearm using discrete skin-slip. A linear array of motors are controlled such that the tactors apply a pre-determined skin-contact profile to the forearm. \review{The number of tactors, N, is 5 and the distance between tactors, D, is 20~mm.}}\label{fig:rollersetup}
	%\vspace{-0.1in}
\end{figure}

This paper has two main contributions. First, we present the design of a novel haptic device for creating a stroking sensation on the arm using discrete lateral skin-slip. The device, shown in Fig.~\ref{fig:rollersetup}, is comprised of a linear array of motors with a tactor, which sequentially provide skin-slip along the arm. Second, we identify device actuation signal parameters that result in continuous and pleasant sensations through human-subject studies. 
%a human-subject study.
The paper is organized as follows. In Section~\ref{sec:background} we discuss some of the parameters involved in the perception of skin-slip, as well as previous devices that use either physical motion or haptic illusions to create a stroking sensation. Section~\ref{sec:prototype} presents the design and control of our sequential skin-slip device, and Section~\ref{sec:study} evaluates the continuity and pleasantness of the stroking sensations created by the device in a human-subject study. \review{In Section~\ref{sec:paradigm}, we discuss an open response experiment conducted to determine how users describe the sensation and confirm that the sensation is continuous and pleasant without priming the subjects. In Section~\ref{sec:distance}, we present another human-subject study} to understand the effect of spacing between skin contact points on the perceived sensation. Finally, Section~\ref{sec:conclusion} concludes with a summary of findings and need for future work.

\section{Background}
\label{sec:background}
%\hmc{Discussion of perception of skin-slip. What are the mechanoreceptors involved? Dependence on size of tactors or speed? How does this motivate our design?}

Understanding how humans sense lateral motion on the skin is the first step toward creating a haptic device that can realistically mimic these complex sensations. This understanding of perception is even more important when designing a haptic illusion to fool the sense of touch, as we present in this paper. Humans sense touch through specialized cells embedded in the skin called mechanoreceptors, which each sense and respond to a specific type of haptic stimulus. \secondreview{The mechanoreceptors in glabrous (non-hairy) skin include the Pacinian corpuscles, Merkel disks, Ruffini endings, and Meissner corpuscles. These mechanoreceptors respond to various stimuli, such as vibration, skin stretch, and skin-slip~\cite{johnson2001roles}. A stroking sensation involves the combination of these stimuli, with small impact vibrations due to contact with the skin, and skin deformation such as normal force, skin stretch, and skin slip occurring with the movement along the skin. Thus, all of these mechanoreceptors are likely stimulated in different ways during stroking. In addition, research has shown that C tactile (CT) afferents exist in hairy skin and help to produce pleasant sensations~\cite{Olausson2002}. CT afferents respond optimally to gentle stroking touch~\cite{mcglone2014discriminative} and respond maximally to stroking in the range of 1-10 cm/s, which has also been shown to be the most pleasant range of speeds for stroking on the skin~\cite{ackerley2014touch}.}
%It is known that Meissner corpuscles respond to skin deformation such as skin-slip~\cite{johnson2001roles}. \review{However, the presence and distribution of various mechanoreceptors differ in hairy and non-hairy skin. %%Research has shown that Meissner corpuscles do not exist in hairy skin~\cite{ackerley2014touch}; instead there exists a different mechanoreceptor, the C tactile (CT) afferent~\cite{mcglone2014discriminative}.
%Specifically, research has shown that C tactile (CT) afferents exist in hairy skin and selectively respond to stroking motions~\cite{mcglone2014discriminative}. Additionally, CT afferents }
%% The CT afferents selectively respond to stroking motions~\cite{mcglone2014discriminative} and
%respond maximally to stroking in the range of 1-10~cm/s, which has also been shown to be the most pleasant range of \review{speeds} %%velocities
%for stroking on the skin~\cite{ackerley2014touch}. 
CT afferents can respond to indentation forces in the range of 0.3-2.5~mN~\cite{Vallbo1999CTafferents}\review{. Biggs and Srinivasan showed that users can consistently identify tangential and normal displacements at the forearm at skin indentation depths of 1.5~mm~\cite{Biggs2002resolution}. }% and indentations of 1.5~mm~\cite{Biggs2002resolution}.
We designed our haptic device such that the discrete lateral skin-slip will be driven by parameters that address these characteristics of CT afferents \review{and can be easily detected by the user}.

%\hmc{Paragraph on devices that use physical motion to create a stroke.}

Several haptic devices have previously been created to display a stroking sensation using a range of different modalities of haptic stimulation using physical motion. One research group has explored directly stimulating the skin with lateral motion provided by a servo motor~\cite{eichhorn2008stroking}, and another has used parallel bars controlled to create lateral movement with shape memory alloy (SMA) actuators~\cite{knoop2015tickler}. Unfortunately, the stroke lengths for each of these techniques is extremely short, 1~cm and 1~mm respectively. Slightly more abstract but still relying on physical motion, a group of researchers has created a stroking sensation via indirect contact with the skin using an air jet~\cite{tsalamlal2015haptic}. The desire to increase the overall stroke length of a sensation and reduce the complexity of the mechanical design serves as strong motivation for the use of haptic illusions instead of relying on mechanical lateral motion, as in these devices.

%\hmc{Paragraph on devices that use haptic illusions to create a stroke.}

Due to the limitations of physical motion created by mechanical devices, researchers have investigated haptic illusions to create a stroking sensation in hairy skin. Likely inspired by the concept of sensory saltation~\cite{Geldard1972sensorysaltation}, the illusion of motion has been created with vibration~\cite{israr2011tactile,Raisamo2013Comparison}, which has then in turn been used to simulate a stroking sensation for social touch applications~\cite{huisman2016simulating}. Although the CT afferents are not as well understood as the mechanoreceptors in glabrous skin, previous researchers have shown that vibrations~\cite{huisman2016simulating,seifi2013first}, air puffs~\cite{tsalamlal2015haptic}, and thermal displays~\cite{wilson2016hot} can be used to elicit a response, even though these modalities do not directly stimulate the CT afferents via a stroking sensation~\cite{liljencrantz2014tactile}. Our previous work attempts to elicit a response from the CT afferents and create a pleasant stroking sensation using only normal indentation~\cite{VCstudy}. While we cannot confirm that we are activating CT afferents without using microneurography, we successfully created a \secondreview{pleasant }stroking sensation with speeds slightly above the range of speeds known to stimulate the CT afferents. Given our prior success using only normal indentation, we believe that we can create an even stronger haptic illusion of a continuous linear motion using discrete lateral skin-slip, as it combines the use of direct lateral motion and illusory techniques.

\section{Device Design}
\label{sec:prototype}
This section describes the design and actuation of a world-grounded haptic device that creates continuous linear sensations along the arm using discrete lateral skin-slip. The focus of our design was to gain an understanding of the skin-slip parameters required to create a continuous and pleasant sensation. Here we discuss the mechanical and electronic design of the device and the actuator command signals.

\begin{figure}[t!]
	\centering	\includegraphics[width=0.9\columnwidth]{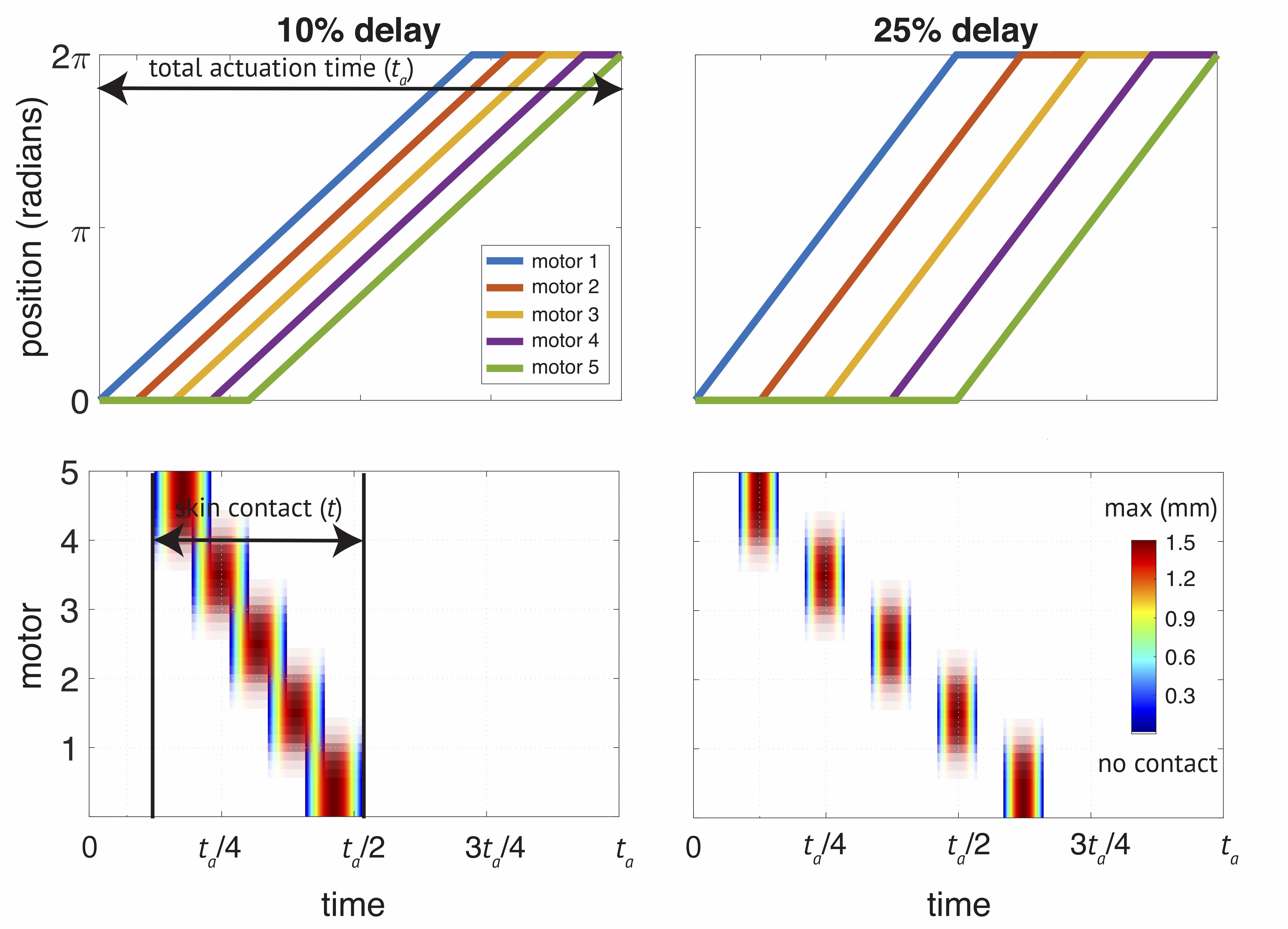}
	\vspace{-0.1in}
	\caption{(Top) Sequential command signals for each motor for completion of a full rotation for 10\% delay and 25\% delay, \secondreview{where $N$ = 5, $D$ = 20~mm, and $\omega$ = 2$\pi$~rad/sec}. (Bottom) Profile of tactor contact on skin over time. Shorter delays result in more overlap of tactor skin contact, while longer delays result in more discrete contacts. %\hmc{I think we should refer to these signals as the "command signals" rather than control signals. The signals that come out of the PID controller would be the control signals. This should be changed in the text as well.}
    }\label{fig:rotationGraph}
	\vspace{-0.15in}
\end{figure}

\subsection{Electro-Mechanical Hardware}

To identify skin-slip patterns that create a continuous and pleasant sensation, we created a system using a linear array of motors (Fig.~\ref{fig:rollersetup}) that apply discrete lateral skin-slip to the forearm. Our device contains an array of five Faulhaber 1624E0175 DC motors, each with a quadrature encoder. The motors have a 141:1 gear ratio, which limits the speed of the motors to 92~RPM, while increasing resolution and torque. %With our system, we are able to control the array of motors so that the end effectors contact the skin at speeds ranging from 2.3 to 24.4~cm/s. This speed is dependent on the amount of time it takes for the tactor to complete a full rotation, which we refer to as the duration of rotation, and amount of delay between the onset of each tactor as a percentage of the rotation duration. \cmn{ADD EQUATION AND EXPLANATION FOR ACTUATOR SPEEDS HERE. Also worth discussing the 1.5~mm of indentation here? ~\cite{Biggs2002resolution}} \srw{This information will be added by Sophia.} \hmc{I think this speed information should be moved to the Actuation Signals subsection. The equations won't make sense if you haven't explained how the actuators are controlled yet.} This range of speeds matches the motions that maximally stimulate the CT afferents (mechanoreceptors in hairy skin that respond to stroking sensations)~\cite{mcglone2014discriminative,ackerley2014touch}.

We attached a rounded tactor to each motor shaft (Fig.~\ref{fig:rollersetup}). The tactor is the element of the device that contacts the skin. It is mounted on the motor shaft using a coupler that is press-fit directly onto the shaft of the motor. The other side of the coupler has a + shaped cross-section that prevents the tactor from rotating due to the torque produced by contact with the skin. We iterated through several different tactor designs, varying the roundedness of the tactor edge and the material adhered to the end-effector of the tactor, including silicone and Dycem. After pilot testing directly comparing these different designs, we decided to laser-cut the tactors from 1/4-inch acrylic because this material created the most pleasant sensation while ensuring that the interaction produced \review{primarily }%only
skin-slip. % and not skin-stretch. 
\review{Yem et al. use rotational motion of rigid ball effectors in a similar fashion to provide skin-slip to the wrist for directional cues~\cite{Yem2015}.}
\review{Our setup allows us to create uniform tactor elements and therefore consistency in the applied sensation. Thus, we conduct an experiment in which we can investigate the performance of the haptic illusion. As briefly described in~\cite{VCstudy}, we previously created haptic sketches to identify methods for creating a salient, continuous, and pleasant stroking sensation. This process of haptic sketching included several soft materials, but we found that a series of rigid contacts could be used to generate the desired sensations. While our current design uses rigid contacts, future research could be performed to explore the sensations created by softer materials, like brushes or other compliant materials.}
The motors are mounted in 3-D printed motor holders, or carriages, to firmly fix the round motors in place between two independently adjustable stands, which hold the forearm in place. The stands allow us to align the position of the elbow and the wrist for consistent indentation of tactors 1.5~mm into the user's skin.

\subsection{Actuation Signals}
\label{subsec: actuation}
%The linear array of tactors creates the sensation of a stroke along the arm by sequentially providing lateral skin-slip.

The tactors individually create a short skin-slip sensation on the arm by rotating the motors over a short path. \secondreview{The tactors begin off of the skin, rotate to first make normal contact with the skin, and then stretch/slide along the skin until they slip off the skin.} The motors are actuated one at a time to create a set of sequential skin-slips along the arm, which together create a longer stroking sensation. The feeling of this stroke can be controlled by varying the rotation speed of the tactors \secondreview{(angular velocity)} and the amount of delay between the onset of rotation for adjacent tactors. A control system, implemented in C++, sets the trajectories of the tactors. The software reads encoder values from the motors and implements a PID controller to set the position of the tactor. The motors are driven using a Sensoray 826 PCI card at 10~kHz via a linear current amplifier. The linear current amplifier was constructed using a power op-amp (LM675T) with a gain of 1~A/V. The Faulhaber motors are rated to a peak current of 10~mA. The described current amplifier circuit provides the necessary current for the motors at the desired voltages without exceeding the maximum output current of the Sensoray board.

The tactors in the array are sequentially activated using the same signal with a set \secondreview{angular velocity} %rotation speed
and amount of delay between the onset of rotation for adjacent actuators. The effect of this delay can be seen in Fig.~\ref{fig:rotationGraph}. The signals on the left are delayed by 10\% of the amount of time to complete a full rotation, which results in overlapping skin contact. The signals on the right are delayed by 25\% of the amount of time to complete a full rotation, which causes no overlapping skin contact. For each \secondreview{angular velocity}%duration of tactor rotation
, shorter delays result in more overlap of tactor skin contact and longer delays result in more discrete skin contact. We study the effects of this delay between tactors and the \secondreview{angular velocity} %varying speeds of rotation
(both local and apparent \review{speed}) %velocity
on the perceived continuity and pleasantness of the stroke in Section~\ref{sec:study}.

\begin{figure}[!t]
	\centering
	\includegraphics[width=0.45\columnwidth]{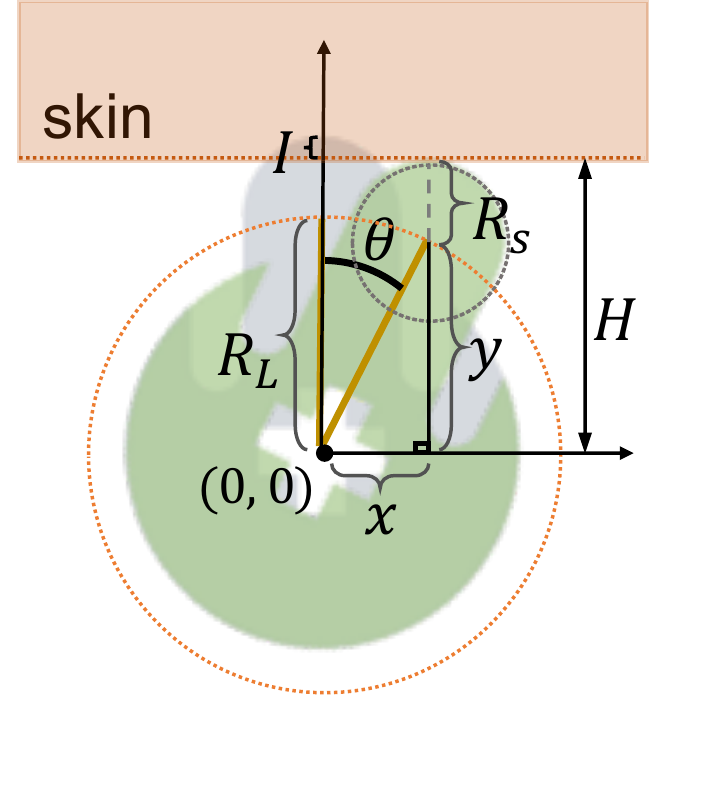}
	\vspace{-0.35in}
	\caption{\secondreview{Illustration of a tactor's indentation into the skin, $I$, and movement across the surface of the skin. The configuration of the tactor during maximum indentation is shown in gray. The configuration of the tactor just as it is leaving the skin is shown in green.}} \label{fig:contactFig}
	\vspace{-0.2in}
\end{figure}

The CT afferents respond optimally to speeds in the range of 1-10 cm/s~\cite{mcglone2014discriminative,ackerley2014touch}. Thus, we quantified the speed of our device on the skin to evaluate how efficient our device is at stimulating the CT afferents. The speed is calculated using the distance that the tactor travels along the skin, which is dependent on several variables, illustrated in Fig.~\ref{fig:contactFig}. These dependencies include the radius of the rounded tactor tip that is in contact with the skin, $R_s$, the radius of the trajectory from the center point to the rounded top, $R_L$, and the distance from the center of rotation to the skin, $H$. The vertical contribution of the movement, $y$, is described by the following equation:
\begin{equation}
y = R_L \cos(\theta)
\end{equation}
\secondreview{When the tactor is in contact with the skin, such that
\begin{equation}
    y+R_S \geq H,
\end{equation}
the amount of tactor indentation into the skin is:
\begin{equation}
I = I_{max} - (R_S + R_L) - (y + R_S)
\end{equation}}
When the tactor is leaving the skin, \secondreview{which is the exact configuration shown in green in Fig.~\ref{fig:contactFig} and when $I$ = 0,} the following condition must hold:
\begin{equation}
H = y + R_S
\end{equation}
Given this constraint, we can calculate the corresponding horizontal contribution, $x$ (Eq. \ref{eq:x}), and the associated angle $\theta$ (Eq. \ref{eq:theta}).

\begin{equation} \label{eq:x}
x = R_L \sin \left(\arccos\left( \frac{H-R_S}{R_L}\right) \right)
\end{equation}

\begin{equation} \label{eq:theta}
\theta = \arccos\left(\frac{H-R_S}{ R_L}\right)
\end{equation}

The total movement of the tactor along the skin is, consequently, equal to $2x$. For our tactor specifications, \review{we used $R_s$ = 3~mm, $R_L$ = 9~mm, and \secondreview{$I_{max}$} = 1.5~mm, 
%$I$} = 1.5~mm
this }%that
means that one actuator travels 1.0 cm along the skin. The total time, $t$, that the tactors travel along the skin is dependent on \secondreview{the angular velocity of the tactor, $\omega$,}
%the amount of time for the tactor to complete a full rotation \review{duration of rotation, $T$,}
the delay, $d$, and the number of tactors, $N$, is: 
\begin{equation}
 \secondreview{t =\left(\frac{2\pi}{\omega}\right)\left(\frac{\theta}{\pi} + d (N - 1)\right)}
%t = \frac{\theta}{\pi}\left(\review{T}\right) + \review{T} d (N - 1)
\end{equation}

\secondreview{While $t$ is the total time that the tactors travel along the skin, the tactors are actuated for a longer period of time, because the tactors must complete the rotation to return to their initial position. This total actuation time, $t_a$, is:}
\secondreview{
\begin{equation} \label{eq:t_a}
t_a = \left(\frac{2\pi}{\omega}\right) (1 + d (N - 1))
\end{equation}}
The speed of the device is defined by two parameters, the local speed of the tactor, $v_{local}$, which is the speed of the tactor as it slips along the skin, and the apparent speed, $v_{apparent}$ of the lateral motion, which is the average speed of the contact traveling along the arm. The local speed is given by Eq.~\ref{eq:local1}\finalreview{ and Eq.~\ref{eq:local2} } \secondreview{and is based on the total time, $t$, it takes for one tactor ($N$ = 1) to move along the skin}. The apparent speed is related to the distance between the tactors, $D$, \secondreview{and the total time, $t$, that $N$ tactors travel along the skin} (Eq.~\ref{eq:apparent}).
\begin{equation} \label{eq:local1}
v_{local} = \frac{2x}{t} \qquad \mathrm{where ~} N = 1 \finalreview{\mathrm{,}}
\end{equation}
\finalreview{
\begin{equation} \label{eq:local2}
\finalreview{\mathrm{therefore} \qquad v_{local} = \frac{x\omega}{\theta}}
\end{equation}}
\begin{equation} \label{eq:apparent}
v_{apparent} = \frac{2x + D(N-1)}{t}
\end{equation}

The apparent speeds for all combinations of delay and duration of rotation are presented in in Table~\ref{table:apparent}, along with the local speeds for all conditions (which only vary based on \secondreview{angular velocity} %duration of rotation
and is not dependent on delay). %We conducted internal pilot studies to determine which delay and pulse duration parameters should be investigated during the study. We initially piloted illusory strokes with slower speeds closer to 1~cm/s, but they did not elicit a sensation similar to a pleasant stroke and so they were not included in the investigation.

The apparent \review{speed} %velocity
for each \secondreview{angular velocity} %duration of rotation
is always larger than the local \review{speed}%velocity
. This is because at delays larger than 30\% the tactors trajectories begin overlapping with each other. This could be remedied by either making the distance between tactors larger or decreasing the length of the tactor. \review{Initially }we chose to minimize the distance between tactors as we hypothesized that it would create the strongest illusion. While we could have designed shorter tactor tips, it would bring the forearm closer to the motor shafts. Decreasing the length in the tactor increases the chance that the forearm could contact the motors which would obscure the sensation of the skin-slip.

%\begin{table}
%\caption{Apparent Speed (cm/s)}
%\begin{tabular}[b]{|c|c|c|c|c|c|}
%\hline
%	 & \textbf{1000 ms} & \textbf{1500 ms} & \textbf{2000 ms} & \textbf{2500 ms} & \textbf{3000 ms} \\ \hline
%	\textbf{0\%} & 48.2 & 32.2 & 24.1 & 19.3 & 16.1\\ \hline
%	\textbf{5\%} & 23.3 & 15.5 & 11.6 & 10.7 & 7.8 \\ \hline
%	\textbf{10\%} & 15.3 & 10.2 & 7.7 & 6.1 & 5.1 \\ \hline
%	\textbf{15\%} & 11.4 & 7.6 & 5.7 & 4.6 & 3.8 \\ \hline
%	\textbf{20\%} & 9.1 & 6.1 & 4.6 & 3.6 & 3.0 \\ \hline
%	\textbf{25\%} & 7.6 & 5.1 & 3.8 & 3.0 & 2.5 \\ \hline
%	\textbf{30\%} & 6.5 & 4.3 & 3.2 & 2.6 & 2.2 \\ \hline
%\end{tabular} \label{table:apparent}
%\end{table}

%\begin{table}
%\caption{Local Speed (cm/s)}
%\begin{center}
%\begin{tabular}[b]{|c|c|c|c|c|}
%\hline
%	\textbf{1000 ms} & \textbf{1500 ms} & \textbf{2000 ms} & \textbf{2500 ms} & \textbf{3000 ms} \\ \hline
%	 5.3 & 3.6 & 2.7 & 2.1 & 1.8\\ \hline
%\end{tabular} \label{table:local}
%\end{center}
%\end{table}

\section{User Study \review{to Understand Actuation Parameters}}
\label{sec:study}
To identify actuation parameters that create a continuous, pleasant sensation, we ran a study with 16 participants (14 right-handed, 2 ambidextrous; 11 male, 5 female; aged 20-48). Ten participants were very familiar with haptic devices and six were not. The protocol was approved by the Stanford University Institutional Review Board, and all participants gave informed consent.

\subsection{Methods}

\begin{figure}[b]
	\centering
		\vspace{-0.2in}
	\includegraphics[width=0.4\columnwidth]{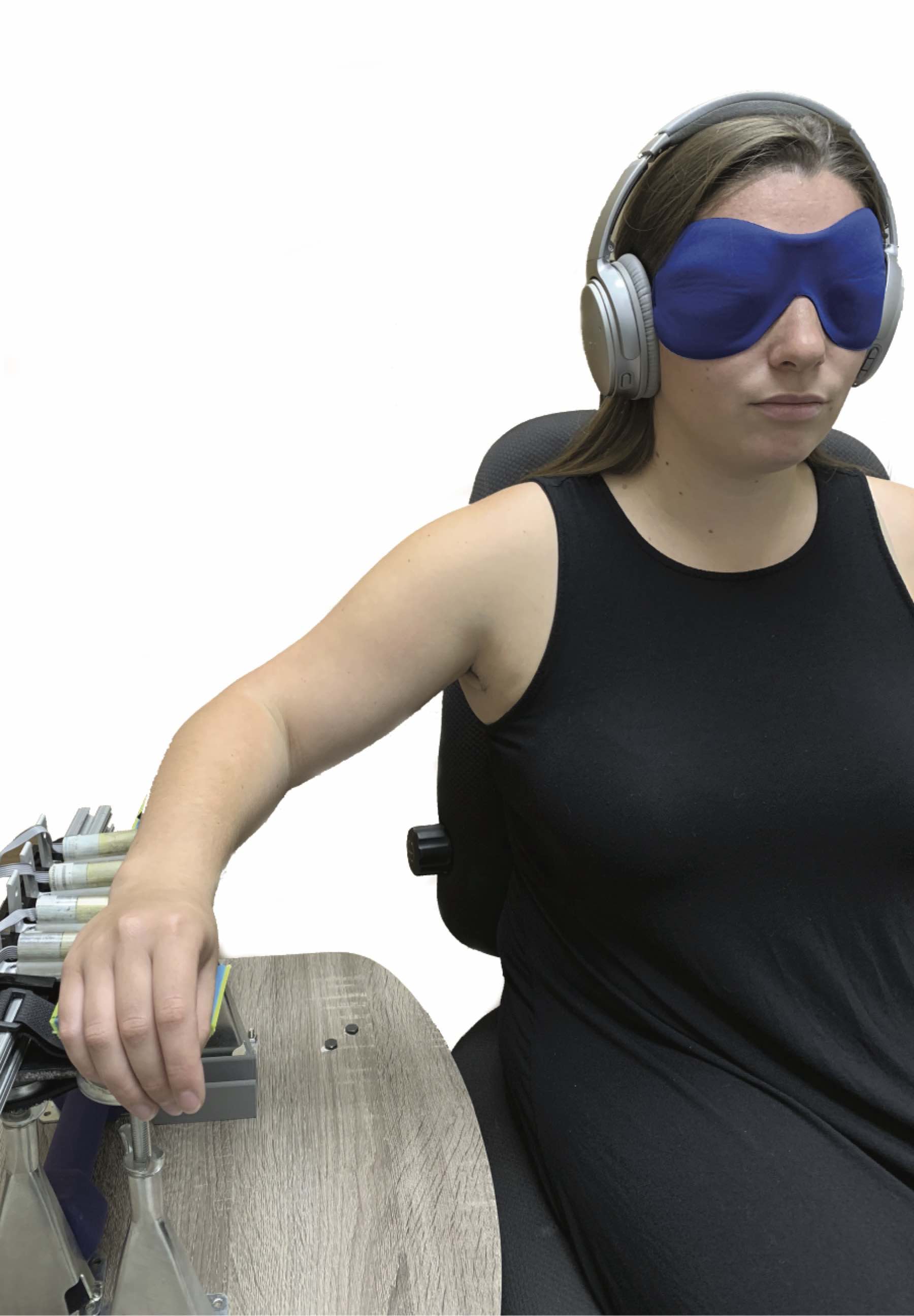}
	\vspace{-0.1in}
	\caption{The setup for the human subjects studies. Participants wore noise canceling headphones placed their right volar or dorsal forearm into the device. In the open response study \review{and contact spacing study}, subjects only placed their right volar forearm into the device and were also required to wear a blindfold.}\label{fig:studysetup}
	%\vspace{-0.2in}
\end{figure}

\begin{figure}[!t]
	\centering
	\includegraphics[width=0.7\columnwidth]{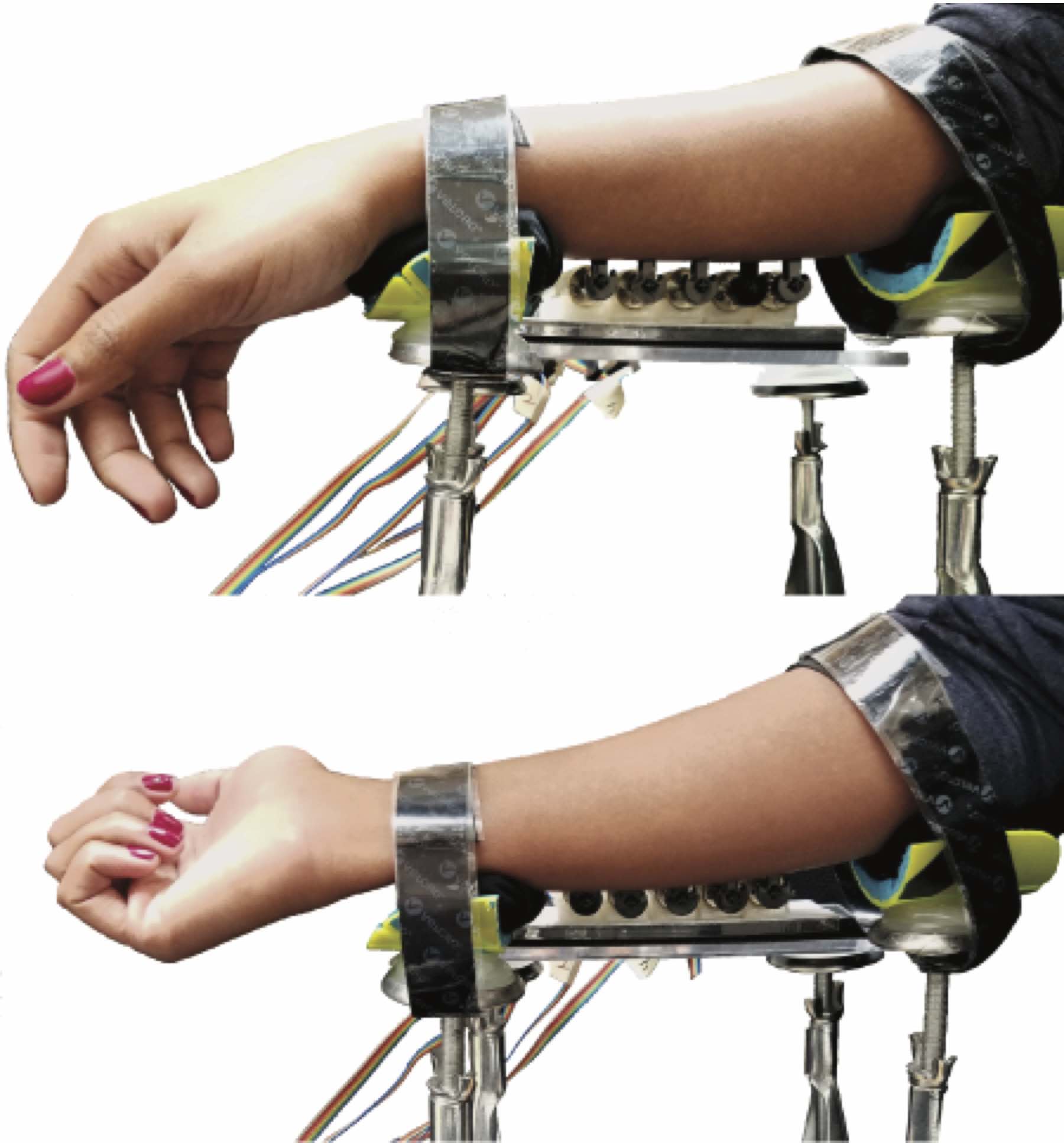}
	\vspace{-0.1in}
	\caption{(Top) User with their volar forearm placed in the haptic device. (Bottom) User with their dorsal forearm placed in the haptic device.}\label{fig:userstudysetup}
	\vspace{-0.1in}
\end{figure}

\label{sec:methods}
Participants sat at a table and placed their right wrist and elbow onto the haptic device. The participants had their arm at their side and faced forward, so they were unable to see the motors and tactors (Fig.~\ref{fig:studysetup}). The participants wore headphones playing white noise to block sounds produced by the motors. \review{Participants heard white noise both during the trials and the intervals between trials.}

Participants completed the study with two contact locations shown in Fig.~\ref{fig:userstudysetup}: (1) with the tactors contacting the underside (volar side) of their forearm and (2) with the tactors contacting the top side (dorsal side) of their forearm. \finalreview{We chose to investigate the volar and dorsal forearm because both are convenient locations for body-mounted wearable devices. Although they are both classified as hairy skin, we hypothesize that the density of mechanoreceptors at these locations differ and that will influence the perceived sensation. }The first contact location for each participant was pseudo-random and balanced across participants to mitigate order effects. Before beginning the study, we aligned the participant's elbow and wrist to ensure the tactors would indent 1.5~mm into the skin \secondreview{($I_{max}$ = 1.5~mm)}. 
%\review{($I$ = 1.5~mm)}.
\review{We then rotated the tactors so that they would not be in contact with the skin at the initial phase of each trial. Each tactor started at the negative 90-degree position (where 0 degrees is defined as being perpendicular and indented into the user’s skin).}

In the study, we varied the \secondreview{angular velocity (2$\pi$, 1.33$\pi$, $\pi$, 0.8$\pi$, and 0.66$\pi$~rad/s)} %duration of tactor rotation (1.0, 1.5, 2.0, 2.5, and 3.0 seconds) 
and the amount of delay between actuators (0\%, 5\%, 10\%, 15\%, 20\%, and 25\% of rotation duration) so that we could understand the effects of both local and apparent \review{speed} %velocity 
(Table~\ref{table:apparent}). \review{The number of tactors, $N$ = 5, and the distance between contact points, $D$ = 20~mm, stayed constant throughout all of the trials.} This resulted in 30 unique actuation conditions, each of which was displayed twice. The order of conditions was randomized, and participants completed all 60 trials for one forearm location before switching to the next location. Each participant completed a total of 120 trials broken into 4 blocks of 30 trials. Between each of the blocks, participants were given a 2 minute break and the tactors were realigned. \review{The participants were allowed to remove the headphones playing white noise during this break.} %Participants were allowed to repeat the sensation as many times as desired.
On average, participants completed the study in under 1 hour.

After feeling each condition, participants rated the sensation on its perceived continuity and pleasantness. Participants rated continuity using a 7-point Likert scale (1=Discrete and 7=Continuous). Similarly, they rated pleasantness on a Likert scale ranging from -7 to +7 (-7=Very Unpleasant, 0=Neutral, +7=Very Pleasant). After completing all 120 trials, participants completed a post-study survey which asked participants to rate using a 7-point Likert scale how difficult it was to distinguish sensations between trials, whether it was easier to distinguish between sensations on the volar or dorsal forearm, and if the sensations felt stronger on the volar or dorsal forearm and were also given space to provide any additional comments.

\begin{table}[t]
	\centering
    \caption{Computed Apparent Speeds of Contact Point \review{($N$ = 5 and $D$ = 20~mm)}}
    	\vspace{-0.1in}
	\includegraphics[width=\columnwidth]{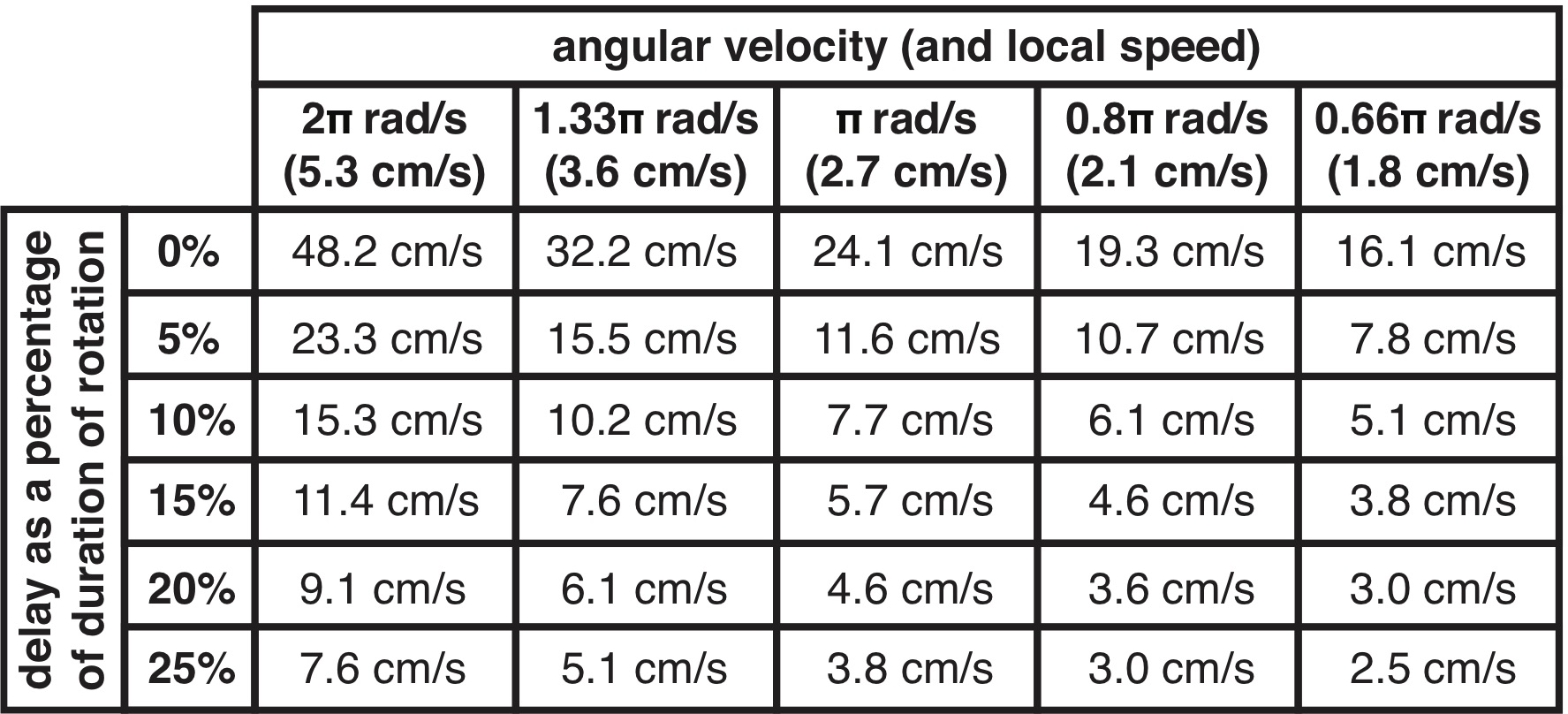}
%	\vspace{-0.2in}
%	Computed apparent speeds of the contact point on the skin are shown for each combination of delay and duration of rotation (local speed in parenthesis) investigated in the study.
    \label{table:apparent}
	\vspace{-0.2in}
\end{table}

\subsection{Results}
\label{results}
\begin{figure*}[t]
	\centering
	\includegraphics[width=1.8\columnwidth]{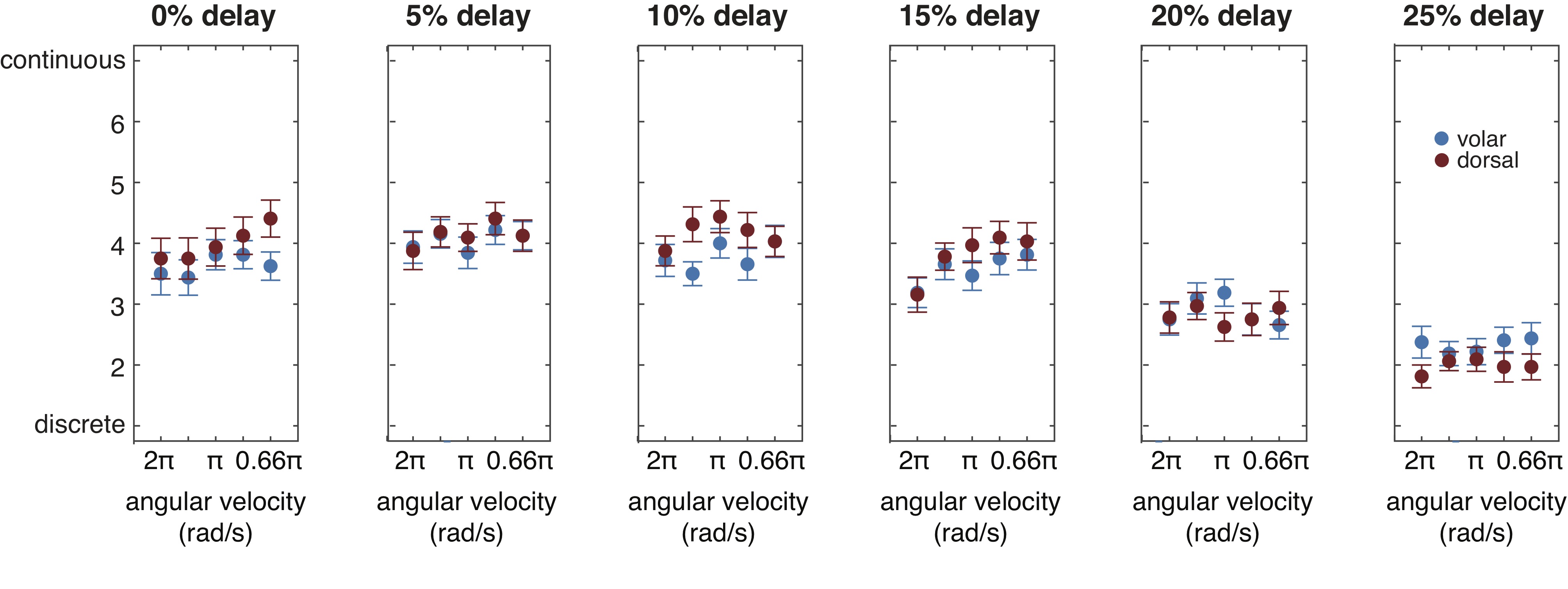}
	\vspace{-0.25in}
	\caption{Average continuity ratings of all participants with standard error bars.} %Linear regression was done on the average ratings ($C$ is the continuity, $T$ is time).}
	\label{fig:contGraph}
	\vspace{-0.1in}
\end{figure*}

\begin{figure*}[t]
	\centering
	\includegraphics[width=1.8\columnwidth]{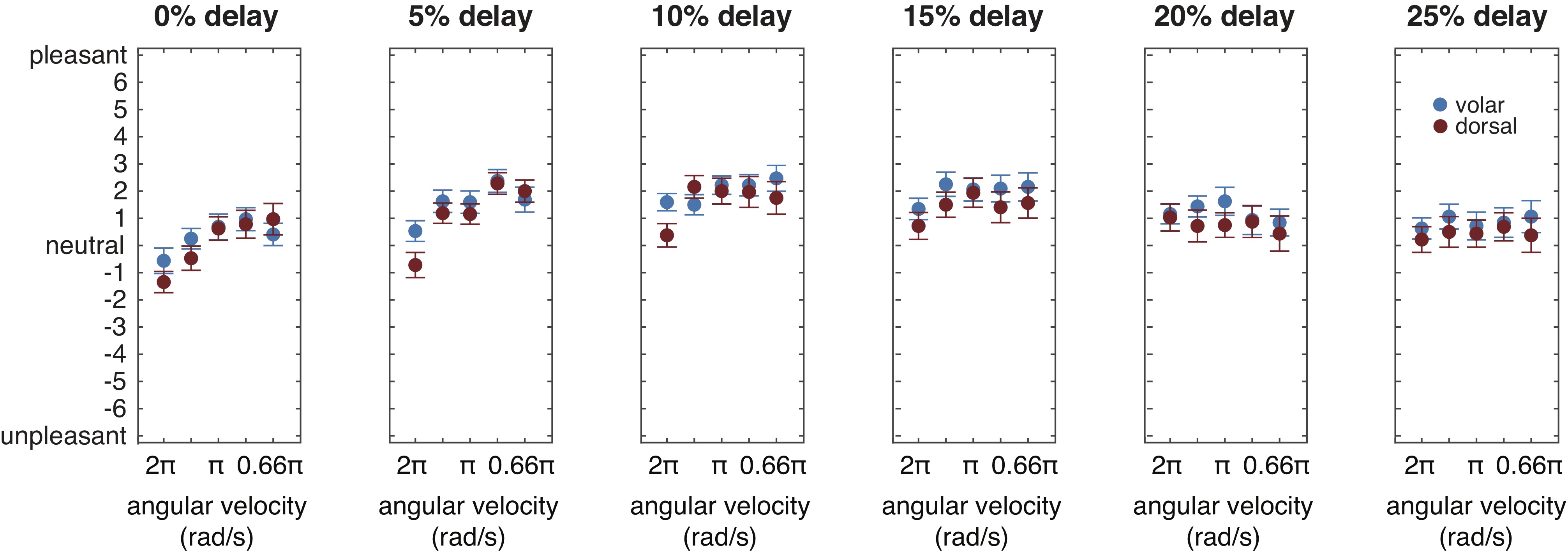}
	\vspace{-0.1in}
	\caption{Average pleasantness ratings of all participants with standard error bars.} %Quadratic fit was done on the average ratings ($P$ is the pleasantness, $T$ is time).}
	\label{fig:pleasGraph}
	\vspace{-0.1in}
\end{figure*}

%\begin{table}[!t]
%	\begin{center}
%    \caption{Average Continuity Ratings}
%        \label{table:Continuity}
%\includegraphics[width=\columnwidth]{Average_Continuity_Ratings_final.eps}
%%	\vspace{-0.2in}
%\end{center}
%\scriptsize{This table includes the average continuity ratings for each combination of duration of rotation (local speed) and delay. The top row for each delay corresponds to the bottom of the forearm (volar) and the bottom row corresponds to the top of the forearm (dorsal).}
%    %\label{table:Continuity}
%%	\vspace{-0.1in}
%\end{table}

%\begin{table}[!t]
%	\begin{center}
%    \caption{Average Pleasantness Ratings}
%        \label{table:Pleasantness}
%	\includegraphics[width=\columnwidth]{Average_Pleasantness_Ratings_final.eps}
%%	\vspace{-0.2in}
%\end{center}
%	\scriptsize{This table includes the average pleasantness ratings for each combination of duration of rotation (local speed) and delay. The top row for each delay corresponds to the bottom of the forearm (volar) and the bottom row corresponds to the top of the forearm (dorsal).}
%    %\label{table:Pleasantness}
%	%\vspace{-0.1in}
%\end{table}

%Table~\ref{table:Continuity} and 
Figure~\ref{fig:contGraph} shows the average continuity rating across all participants, separated by delay and \secondreview{angular velocity. }%duration of rotation.
\secondreview{}
%We fit a linear regression to the average continuity ratings for each delay. The positive slope values of the fitted regressions indicated that rated continuity increases with increasing duration of rotation value when delay is held constant. Similarly, from looking at the decreasing intercepts and slopes of the fitted regressions, it can be seen that continuity decreases with delay.

We ran a three-way repeated measures ANOVA on the continuity ratings with forearm location, delay, and \secondreview{angular velocity }%duration of rotation
as factors. \secondreview{Both delay and angular velocity violated the assumption of sphericity so we used the lower-bound estimate of $\varepsilon$ = 0.25 to correct our calculations. }There was no significant difference in continuity ratings between the volar and dorsal forearm \secondreview{($F(0.25,467.25)=2.66$, $p=0.106$, ${\eta_p}^{2} = 0.001$).}  %\review{($F(1,1869)=2.66$, $p=0.103$, ${\eta_p}^{2} = 0.001$).
The interaction between arm location and delay value was \secondreview{not }significant 
\secondreview{($F(1.25,467.25)=3.05$, $p=0.072$, ${\eta_p}^{2} = 0.008$) 
%($F(5,1869)=3.05$, $p=0.010$, ${\eta_p}^{2} = 0.008$)
and neither was the interaction between forearm location and angular velocity} %while the interactions between forearm location and \secondreview{angular velocity }%duration of rotation
\secondreview{($F(1,467.259)=0.33$, $p=0.566$, ${\eta_p}^{2} = 0.001$).} %($F(4,1869)=0.33$, $p=0.858$, ${\eta_p}^{2} = 0.001$) was not significant.}

Continuity was significantly different across delays \secondreview{($F(1.25,467.25)=90.1$, $p < 0.001$, ${\eta_p}^{2} = 0.194$).} %\review{($F(5,1869)=90.1$, $p=4.26\times10^{-85}$, ${\eta_p}^{2} = 0.194$).
The interaction between delay and \secondreview{angular velocity }%duration of rotation
was not significant \secondreview{($F(5,467.25)=0.7$, $p=0.624$, ${\eta_p}^{2} = 0.007$)}. %($F(20,1869)=0.7$, $p=0.830$, ${\eta_p}^{2} = 0.007$).
To further evaluate the effect of delay, we ran a post-hoc pairwise comparison test with a Bonferroni correction. The results from this test are shown in Table~\ref{table:ContinuityStatsDelay}. To summarize, for smaller delays (0\%, 5\%, 10\%), continuity values were generally not significantly different from the continuity values of the adjacent delays. However, for larger delays, (15\%, 20\%, 25\%), continuity values were generally significantly different from the continuity values of the adjacent delays. This shows that while continuity is strongly linked to the delay between the onset of actuation, small changes in delay do not have a significant effect on the sensation at small delay values. However, beginning at 15\%, small changes in delay have a significant effect at large delay values and the sensation feels less and less continuous.

The results of the ANOVA showed that continuity was \secondreview{not }significantly different across \secondreview{angular velocity }%duration of rotation
\secondreview{($F(1,467.25)=2.66$, $p=0.104$, ${\eta_p}^{2} = 0.006$).}
%\review{($F(4,1869)=2.66$, $p=0.031$, ${\eta_p}^{2} = 0.006$)}.
%%% removed below as a part of the final review (R3 comment that no need to include when no sig effect in ANOVA)
% \secondreview{We ran a post-hoc pairwise comparison test with a Bonferroni correction to confirm this result }%However, we ran a post-hoc pairwise comparison test with a Bonferroni correction to further evaluate the effect of the \secondreview{angular velocity }%duration of rotation
% and found that continuity ratings were not statistically significantly different for any of the pairs of \secondreview{angular velocities }%durations of rotation
% (Table~\ref{table:ContinuityStatsPulse}).
%%% This is a strikethrough version to submit with the revisions
% \finalreview{\sout{We ran a post-hoc pairwise comparison test with a Bonferroni correction to confirm this result and found that continuity ratings were not statistically significantly different for any of the pairs of angular velocities (Table~\ref{table:ContinuityStatsPulse}).}}
These results show that perceived continuity varies and can be controlled by changing the delay of the onset of actuation between motors, regardless of the chosen arm location or \secondreview{angular velocity. }%duration of rotation duration.
Further, these results show that differences in perceived continuity are due to changes in the apparent \review{speed}%velocity
, and not to changes in the local \review{speed}.%velocity.

%Table~\ref{table:Pleasantness} and
Figure~\ref{fig:pleasGraph} shows the average pleasantness rating across all participants, separated by delay and \secondreview{angular velocity. }%duration of rotation.
%We applied a quadratic regression to the data, following previous studies that identified this as an appropriate model for pleasantness~\cite{huisman2016simulating,Loken2009}. The plot indicates that pleasantness is highest for the middle rotation times (1.5, 2.0, and 2.5~s with local speeds of 3.6, 2.7, and 2.1~cm/s, respectively) when delay is held constant. The pleasantness also follows a similar parabolic fit pertaining to delay. Initially the ratings increase with increasing delay and peak at 10\% and 15\% before decreasing.

Similar to our analysis for continuity, we ran a three-way ANOVA on the pleasantness ratings with forearm location, delay, and \secondreview{angular velocity }%duration of rotation
as factors. \secondreview{We found that both delay and angular velocity violated the assumption of sphericity so we used the lower-bound estimate of $\varepsilon$ = 0.25 to correct our calculations. }Unlike our analysis for continuity, this analysis showed that pleasantness ratings were statistically different between the volar and dorsal forearm 
\secondreview{($F(0.25,467.25)=10.56$, $p=0.019$, ${\eta_p}^{2} = 0.006$). %\review{($F(1,1869)=10.56$, $p=0.001$, ${\eta_p}^{2} = 0.006$).
The interactions between forearm location and delay value 
\secondreview{($F(1.25,467.25)=0.12$, $p=0.785$, ${\eta_p}^{2} = 0.0003$) }
%($F(5,1869)=0.12$, $p=0.987$, ${\eta_p}^{2} = 0.0003$)
and between forearm location and \secondreview{angular velocity }%duration of rotation
\secondreview{($F(1,467.25)=0.55$, $p=0.459$, ${\eta_p}^{2} = 0.001$)}
%($F(4,1869)=0.55$, $p=0.701$, ${\eta_p}^{2} = 0.001$)
were not significant.} From Fig.~\ref{fig:pleasGraph}%and table of the results
, we can conclude that the bottom of the forearm is more pleasant than the top of the forearm.

Pleasantness is also statistically different for delay 
\secondreview{($F(1.25,467.25)=17.23$, $p < 0.001$, ${\eta_p}^{2} = 0.044$). }
%\review{($F(5,1869)=17.23$, $p=1.06\times10^{-16}$, ${\eta_p}^{2} = 0.044$).
The interaction between delay and \secondreview{angular velocity }%duration of rotation
was not significant \secondreview{($F(5,467.25)=1.48$, $p=0.195$, ${\eta_p}^{2} = 0.016$)}. 
%($F(20,1869)=1.48$, $p=0.078$, ${\eta_p}^{2} = 0.016$).}
We ran a post-hoc pairwise comparison test with a Bonferroni correction to further evaluate the effect of delay. The results from this test can be seen in Table~\ref{table:PleasantnessStatsDelay}. These results show that the ratings follow a parabolic trend, as the medial values (10\%, 15\%) are not statistically significantly different from each other, but are different from the values on the ends (0\%, 25\%). This corresponds with what we see in the data that the values peak at 10\% and 15\%.

The results of the ANOVA showed that pleasantness was significantly different across \secondreview{angular velocity }%duration of rotation
\secondreview{($F(1,467.25)=9.18$, $p=0.003$, ${\eta_p}^{2} = 0.019$)}.
%\review{($F(4,1869)=9.18$, $p=2.40\times10^{-7}$, ${\eta_p}^{2} = 0.019$)}.
After running a post-hoc pairwise comparison test with a Bonferroni correction (Table~\ref{table:PleasantnessStatsPulse}), we found that the pleasantness ratings for \secondreview{an angular velocity of 2$\pi$~rad/s }%a rotation duration of 1.0~s
(5.3~cm/s) were significantly less than the pleasantness ratings of all of the other \secondreview{angular velocities. }%durations of rotation.
However, the pleasantness values for the other four \secondreview{angular velocities (1.33$\pi$, $\pi$, 0.8$\pi$, and 0.66$\pi$~rad/s }%durations of rotation (1.5, 2.0, 2.5, and 3.0~s
with local speeds of 3.6, 2.7, 2.1, 1.8~cm/s, respectively) were not statistically significantly different from each other.

To determine if the sensations were actually perceived as pleasant, we ran one-sample t-tests on the pleasantness ratings compared to the neutral rating (pleasantness $=0$). Grouping the pleasantness ratings by delay, the delay values of 5\%, 10\%, 15\%, 20\%, and 25\% had ratings that were statistically greater than zero (\secondreview{$p < 0.001$}). This indicates that these conditions were on average rated as pleasant. The pleasantness ratings for the smallest delay value, 0\% delay, was not statistically different from zero ($p=0.12$). We did not find that the pleasantness ratings were significantly less than zero (unpleasant) for any of the delay values. When we grouped the pleasantness ratings by \secondreview{angular velocity, all of the values (2$\pi$, 1.33$\pi$, $\pi$, 0.8$\pi$, and 0.66$\pi$~rad/s with local speeds of 5.3, 3.6, 2.7, 2.1, and 1.8~cm/s, respectively) }%the duration of rotation, all of the values (1.0, 1.5, 2.0, 2.5, and 3.0~s with local speeds of 5.3, 3.6, 2.7, 2.1, and 1.8~cm/s, respectively)
were statistically greater than zero ($p \leq 0.002$). This indicates that these conditions were on average rated as pleasant.

\begin{table}[t]
	\begin{center}
    \caption{Effect of Delay on Continuity}
    	\vspace{-0.1in}
    \label{table:ContinuityStatsDelay}
\includegraphics[width=\columnwidth]{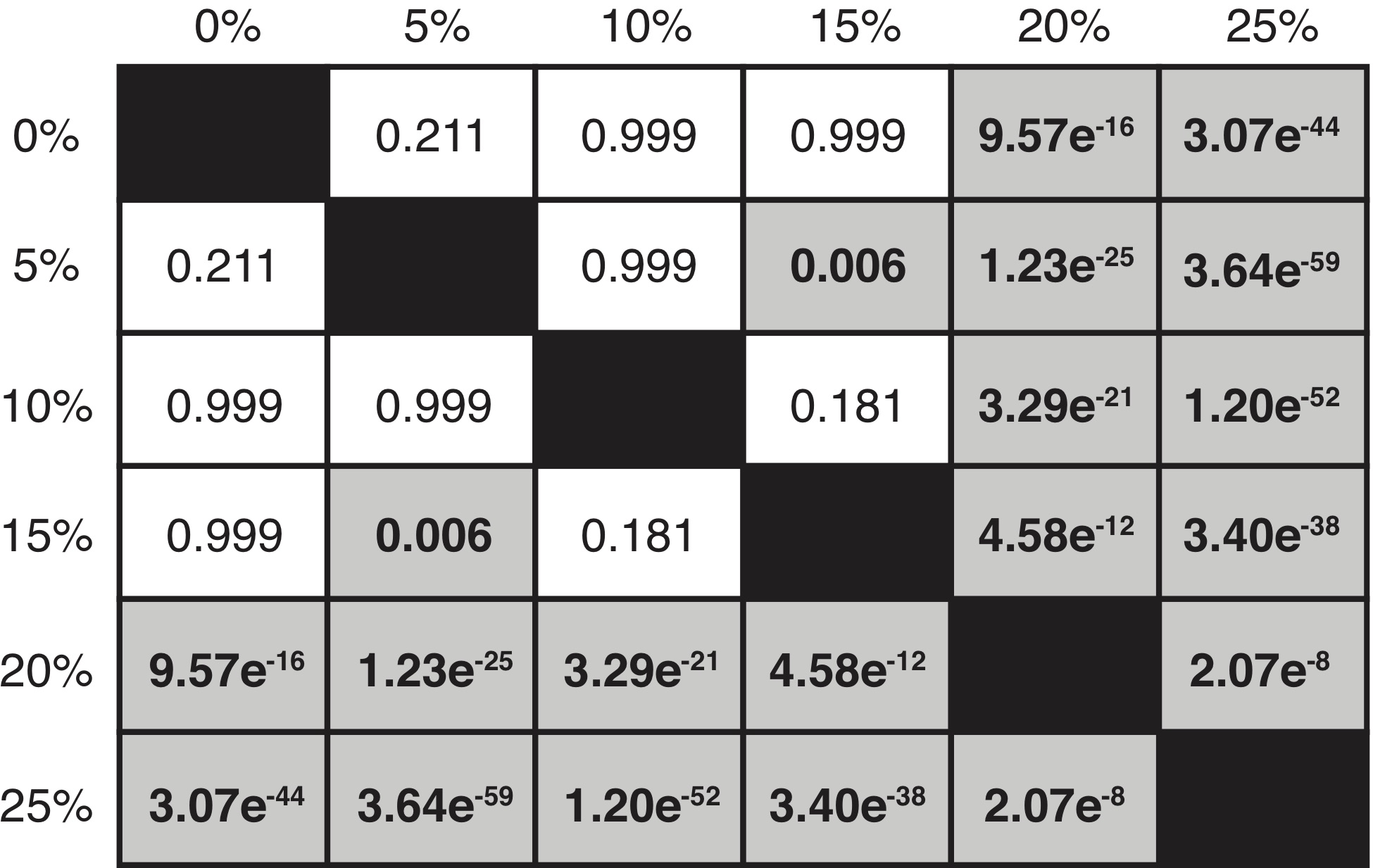}
	%\vspace{-0.2in}
    \end{center}
	\scriptsize{P-values corresponding to the results of the post-hoc pairwise comparison test with a Bonferroni correction for the continuity ratings pertaining to delay. P-values for statistically significant pairs are bolded and shaded in gray.}
    %\label{table:ContinuityStatsDelay}
	\vspace{-0.1in}
\end{table}

%%% removed below as a part of the final review (R3 comment that no need to include when no sig effect in ANOVA)
% \begin{table}[t]
% 	\begin{center}
%     \caption{Effect of Local Speed on Continuity}
%     	\vspace{-0.1in}
%     \label{table:ContinuityStatsPulse}
% 	\includegraphics[width=\columnwidth]{Table_Statistics_Continuity_Pulse_final.eps}
% 	%\vspace{-0.2in}
%     \end{center}
% 	\scriptsize{P-values corresponding to the results of the post-hoc pairwise comparison test with a Bonferroni correction for the continuity ratings pertaining to \secondreview{angular velocity} %duration of rotation
% 	(local speed in parenthesis). There are no statistically significant pairs.}
%     %\label{table:ContinuityStatsPulse}
% 	\vspace{-0.1in}
% \end{table}

%%% This is a strikethrough version to submit with the final revisions
% \begin{table}[t]
% 	\begin{center}
%     \caption{\finalreview{\sout{Effect of Local Speed on Continuity}}}
%     	\vspace{-0.1in}
%     \label{table:ContinuityStatsPulse}
% 	\includegraphics[width=\columnwidth]{Table_Statistics_Continuity_Pulse_final.eps}
% 	%\vspace{-0.2in}
%     \end{center}
% 	\scriptsize{\finalreview{\sout{P-values corresponding to the results of the post-hoc pairwise comparison test with a Bonferroni correction for the continuity ratings pertaining to angular velocity %duration of rotation
% 	(local speed in parenthesis). There are no statistically significant pairs.}}}
%     %\label{table:ContinuityStatsPulse}
% 	\vspace{-0.1in}
% \end{table}

\begin{table}[t]
	\begin{center}
    \caption{Effect of Delay on Pleasantness}
    	\vspace{-0.1in}
    \label{table:PleasantnessStatsDelay}
	\includegraphics[width=\columnwidth]{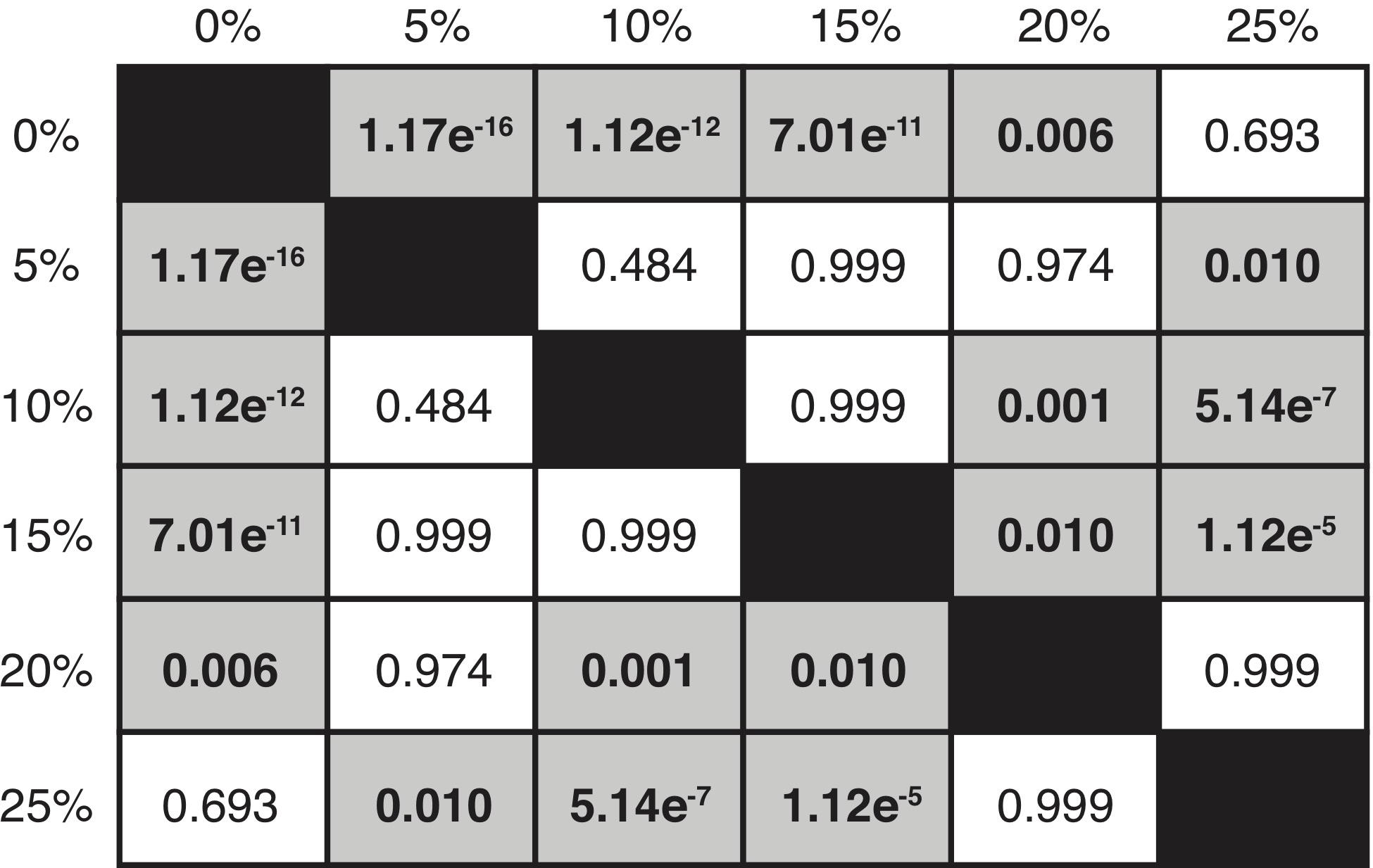}
	%\vspace{-0.in}
    \end{center}
	\scriptsize{P-values corresponding to the results of the post-hoc pairwise comparison test with a Bonferroni correction for the pleasantness ratings pertaining to delay. P-values for statistically significant pairs are bolded and shaded in gray.}
    %\label{table:PleasantnessStatsDelay}
	\vspace{-0.1in}
\end{table}

\begin{table}[t]
	\begin{center}
    \caption{Effect of Local Speed on Pleasantness}
    	\vspace{-0.1in}
    \label{table:PleasantnessStatsPulse}
	\includegraphics[width=\columnwidth]{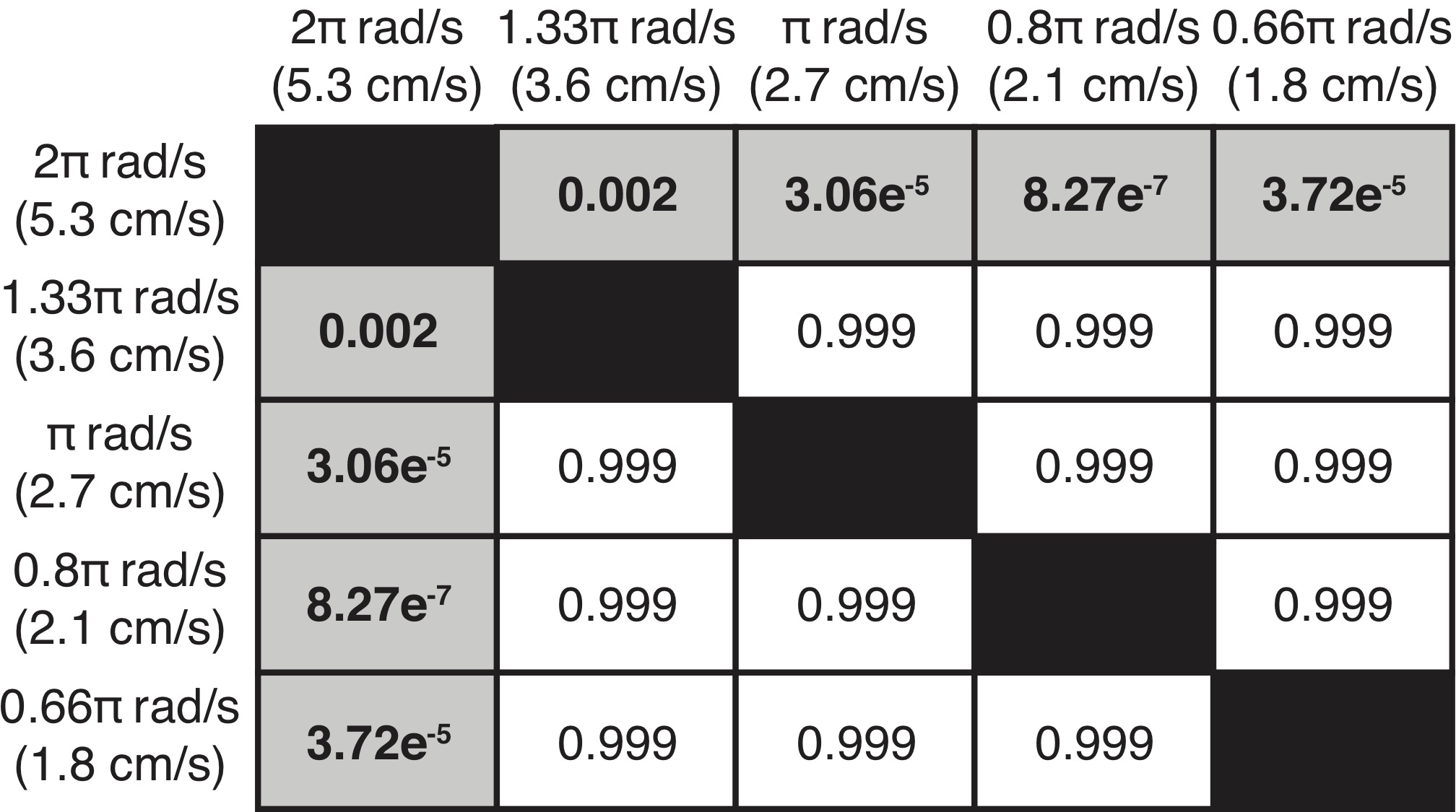}
	%\vspace{-0.2in}
    \end{center}
	\scriptsize{P-values corresponding to the results of the post-hoc pairwise comparison test with a Bonferroni correction for the pleasantness ratings pertaining to \secondreview{angular velocity} %duration of rotation
	(local speed in parenthesis). P-values for statistically significant pairs are bolded and shaded in gray.}
    %\label{table:PleasantnessStatsPulse}
	\vspace{-0.2in}
\end{table}

\subsection{Discussion}
\label{sec:discussion}
This analysis shows how to design signals for the actuation of a discrete lateral skin-slip device. To optimize for a continuous sensation, with our device one should command the motors with minimal delay, such as 5\%, and \secondreview{a slower angular velocity}%maximal duration of rotation
, such as \secondreview{0.66$\pi$~rad/s }%3.0~s
(1.8~cm/s). This combination corresponds to an apparent \review{speed} %velocity
of 7.8~cm/s. To optimize for a pleasant sensation, with our device the motors should be commanded via the medial delay values, either 10\% or 15\%, with a medial \secondreview{angular velocity, such as $\pi$~rad/s }%duration of rotation, such as 2.0~s
(2.7 cm/s). These combinations correspond to apparent \review{speeds} %velocities
of 7.7~cm/s and 5.7~cm/s.

The signal that was rated highest for continuity was 10\% delay with \secondreview{an angular velocity of $\pi$~rad/s }%a duration of 2.0~s
on the dorsal forearm. The effective speed of travel of the sensation along the forearm was 7.7~cm/s, which is within the optimal range of 1-10~cm/s for activating the CT afferents~\cite{ackerley2014touch}. The signal that was rated highest for pleasantness was 10\% delay with \secondreview{an angular velocity of 0.66$\pi$~rad/s }%a duration of 3.0~s
on the volar forearm. The effective speed of travel of the sensation along the forearm was 5.1~cm/s, which is also within the optimal range for activating the CT afferents~\cite{ackerley2014touch}. As previously mentioned, we initially piloted illusory strokes with slower speeds closer to 1~cm/s, but they felt unpleasant. From the parabolic results of our study, we can determine that the perception of touch is more continuous and pleasant when the speed is closer to 10~cm/s than to 1~cm/s. Thus, when creating future haptic devices that involve continuous linear sensations, designers should more specifically focus speeds of 5-10~cm/s for it to be perceived most optimally as continuous and pleasant.

\begin{figure}[b]
	\centering
	\includegraphics[width=\columnwidth]{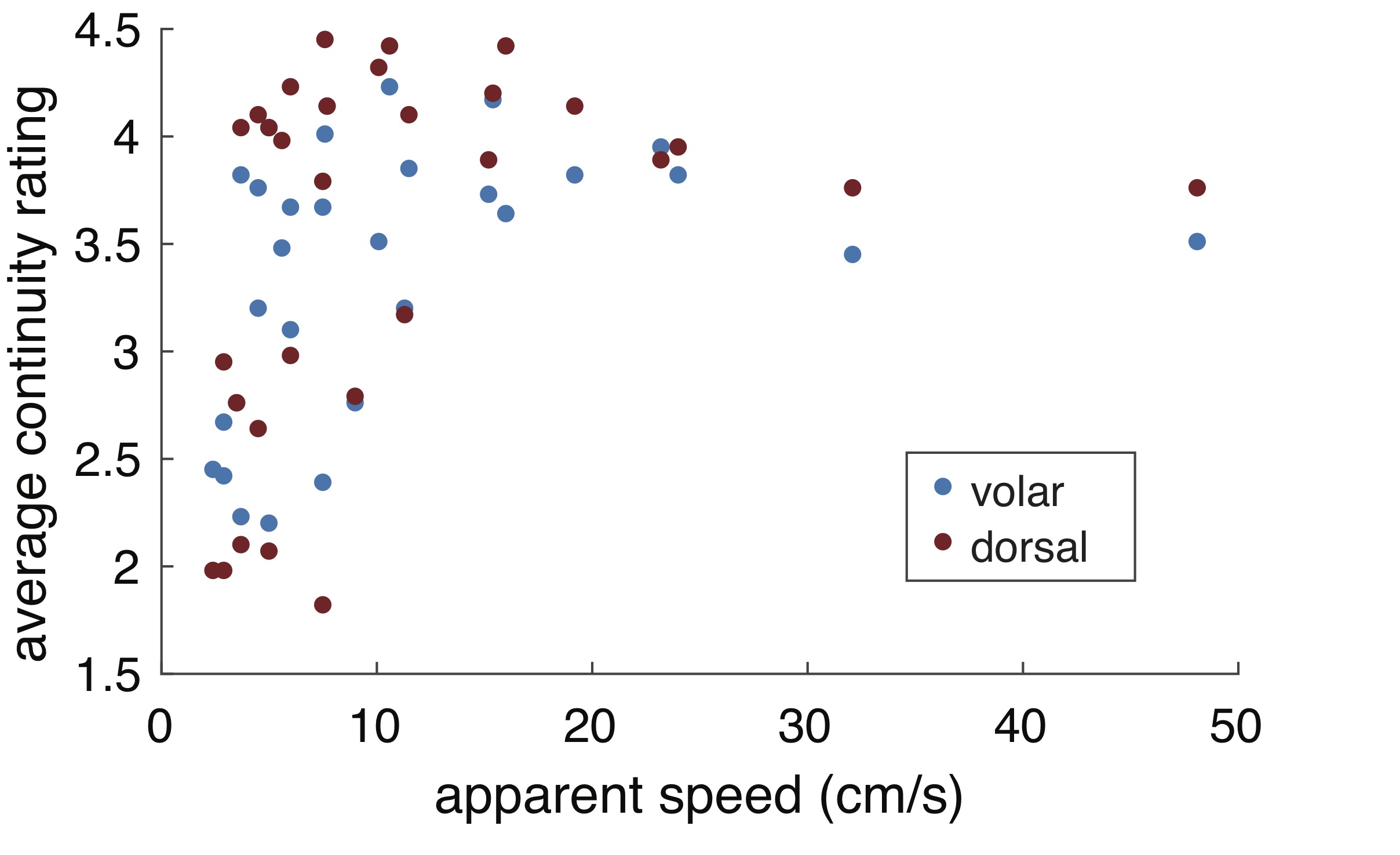}
	\vspace{-0.35in}
	\caption{\review{Average continuity ratings of all participants at each apparent speed for the volar and dorsal forearm.}}\label{fig:contApparentVelocity}
	\vspace{-0.1in}
\end{figure}

\begin{figure}[!t]
	\centering
	\includegraphics[width=\columnwidth]{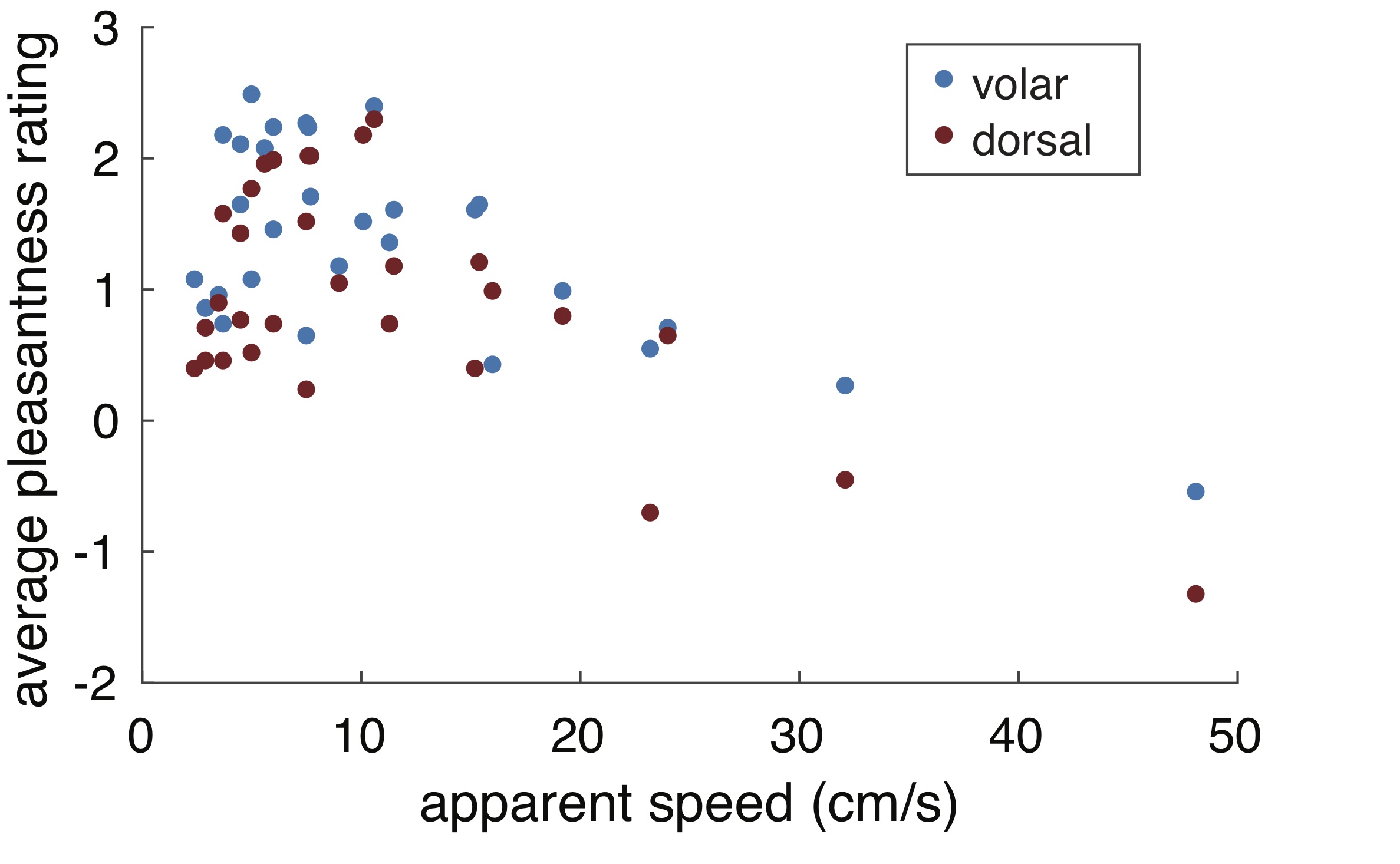}
	\vspace{-0.35in}
	\caption{\review{Average pleasantness ratings of all participants at each apparent speed for the volar and dorsal forearm.}}\label{fig:pleasApparentVelocity}
	\vspace{-0.2in}
\end{figure}

\review{Figures \ref{fig:contApparentVelocity} and \ref{fig:pleasApparentVelocity} show how the average pleasantness and continuity ratings change as a function of apparent speed. Apparent speed is the speed of the illusory motion and is a function of the \secondreview{angular velocity }%duration of rotation
and delay. It is important to note that both delay and \secondreview{angular velocity }%duration of rotation
independently have statistically significant effects on the pleasantness and continuity. However, visualizing apparent speed allows us to see general trends in the data and provides additional knowledge for implementing similar algorithms on other devices. The results show that the average continuity ratings increase approximately linearly as the apparent speed increases until 5-10~cm/s when the ratings begin to decrease slightly before plateauing at higher apparent speeds. This shows that at slower apparent speeds the continuity illusion starts to break down and users, on average, rate the sensation as more discrete. The illusion is the most convincing at apparent speeds between 5 and 15~cm/s. As the apparent speeds continue to increase the sensation of continuity decreases slightly but is still not rated as discrete.  Comparatively, the pleasantness values increase to a peak at 5-10~cm/s before steadily dropping as apparent speed increases. This demonstrates that even though there still exists an illusion of continuity at high apparent speeds, users find the sensation less pleasant. On average, users also find the sensation less pleasant as the apparent speed approaches zero. This decrease follows the same trend as continuity, so it may be possible that the decrease in pleasantness ratings could be a result of the continuity illusion breaking down. It is interesting to note that both the continuity and pleasantness ratings peak at around 5-10~cm/s.} 

Participants' ratings for continuity did not vary significantly between the volar and dorsal forearm. However, their ratings for pleasantness did differ between the two forearm locations. We believe that the reason that there was a difference in pleasantness ratings but not continuity ratings is likely from the design of the device% and not human perception
. Specifically, although we made the device adjustable, it is possible that tactor contact with the skin was different for the two locations due to the different shape and musculature of the forearm. Additionally, in a post-study survey, subjects reported that they preferred to feel the sensation on their volar forearm and felt more comfortable resting their arm in that position than their dorsal forearm. User comfort could be the reason for higher pleasantness ratings for the volar forearm compared to the dorsal forearm. \secondreview{The differences in these ratings could also be due to differing stiffness factors or possibly even different mechanoreceptor densities between the two locations. }%\review{The differences in these ratings could also be due to differing densitites of mechanoreceptors or stiffness factors between the two locations.}
Furthermore, even though the overall pleasantness values were statistically different for the volar and dorsal forearm, the trends across delays and duration of rotation were consistent for the two locations.

Although the average ratings for continuity and pleasantness were not exceptionally high, it is still clear that the device was able to generate a continuous and pleasant linear sensation and that this type of actuation could be successful as part of a wearable haptic device. We believe that because we did not allow the subjects to feel any of the parameters or undergo any training trials prior to the experiment that their responses were not necessarily based on whether or not the sensation felt continuous and pleasant, but how continuous and pleasant the sensation felt in comparison to previous sensations. We also believe that there was inherent variation in the rating methodology between subjects, with some drastically fluctuating from trial to trial and others generally staying close to neutral responses, likely pulled the average values closer to the center. Since our participants were able to see the device before completing the study, we also believe that there is an artificial maximum pertaining to the continuity values. We believe that participants never rated continuity with a 7 because the participants knew in advance that the sensation would not be one continuous motion. In future work, we believe that we could remove this inherent bias by either not allowing the user to see the device in advance of the study or by comparing this sensation to an actual continuous motion (such as a robotic finger dragging along the skin).

\finalreview{As mentioned briefly in the previous paragraph, there was inherent variation in the rating methodology between subjects. This variation between subjects, as well as general human variability, resulted in very small effect sizes (${\eta_p}^{2}$). These small effect sizes do not impact the statistical significance of the results but do highlight that the magnitude of the difference is small. As such, changing the actuation parameters, such as angular velocity and delay, will change the overall sensation, but the change is not drastic.}

We cannot directly compare these results using discrete lateral skin-slip to our previous work using normal indentation~\cite{VCstudy} because we used a different set of participants in the two studies. However, our average continuity and pleasantness ratings are in the same range as those from the normal indentation study. In fact, we had fewer average pleasantness ratings less than 0 and several average pleasantness ratings that were higher than what was collected for the normal indentation study at the same speeds. Therefore, we can confirm that our hypothesis that discrete lateral skin-slip, which combines normal indentation and lateral motion, creates a stronger illusion than normal indentation alone. In future work, we are interested in conducting a study to directly compare normal indentation, discrete lateral skin-slip, and skin stretch to further understand and characterize the parameters of these actuation methods. 

These results provide general models that can be used as guidelines for rendering lateral sensations on the forearm using discrete lateral skin-slip. The results from this study show promise towards creating sensations that could be applicable for social touch (strokes to show comfort or excitement for example), to relay effective directional cues, or other simple messages.

These results also led us to question another important parameter in wearable haptics design: the spacing of skin contact points. Were we only able to successfully create this illusion because our contact points were spaced closely together? Can we spread those contact points further apart and still get similar results? How many contact points are necessary to create the illusion? Can we have fewer contact points and still get similar results? We address some of these questions with \review{an additional user study in Section~\ref{sec:distance}.}
% the open response experiment described next in Section~\ref{sec:paradigm}.

%\section{Open Response Contact Spacing Study}
\section{Open Response \review{Study}}
\label{sec:paradigm}
\review{To qualitatively describe the sensation created by the device and the believability of the continuous skin slip illusion, we conducted an open response user study with 16 subjects (all right-handed; 7 male, 9 female; aged 21-45). Of the 16 subjects, 6 subjects were familiar with haptic devices, but none of the subjects had any previous experience with this haptic device or participated in the previous study. The protocol was approved by the Stanford University Institutional Review Board, and all subjects gave informed consent.}
\subsection{\review{Methods}}
\label{paradigm methods}
\review{Participants were positioned just as in the previous study as it pertains to the volar forearm, described in Section \ref{sec:methods} and shown in Fig.~\ref{fig:studysetup}. The participants did not see the device before the study began and were blindfolded for the entire duration of the study. The participants wore noise cancelling headphones to block out any noise created by the motors. 

Unlike in the previous user study, participants only received the stimulus to the volar side of their forearm. Additionally, while the original design of the haptic device used in the previous user study consisted of an array of five Falhauber motors \review{($N$ = 5)}, we adapted the haptic device to be an array of four motors \review{($N$ = 4)}. This was done such that the device could still fit within the workspace (volar forearm) when increasing the spacing for a contact spacing study (Section~\ref{sec:distance}), which the participants would complete immediately after the open response study. During the open response study the spacing between tactors was set to be $D$ = 20~mm. The haptic signal was played once with \secondreview{an angular velocity of 0.66$\pi$~rad/s }%a duration of 3.0~s
and a delay of 10\%. This \secondreview{angular velocity }%duration of rotation
and delay was chosen because the previous study showed this to be the most pleasant. Participants could ask to repeat the sensation as many times as they desired and were encouraged to repeat until they felt comfortable giving an oral description of the sensation. After feeling the haptic signal, they were asked to describe the sensation they felt and their response was recorded.}

%Unlike in the previous user study, participants only received the stimulus to the volar side of their forearm. The number of tactors used during the open response paradigm was N=4 and the spacing between tactors was set to be $D$ = 20~mm. The haptic signal was played once with a duration of 3.0~s and a delay of 10\%. This duration of rotation and delay was chosen because the previous study showed this to be the most pleasant. Participants could ask to repeat the sensation as many times as they desired and were encouraged to repeat until they felt comfortable giving an oral description of the sensation. After feeling the haptic signal, they were asked to describe the sensation they felt and their response was recorded.}

\subsection{\review{Results}}
\label{paradigm results}
\review{
Five out of 16 participants described the sensations as moving along the arm from the wrist to the forearm. Two explicitly stated that it felt \say{continuous}, while the others implied that it felt continuous via their descriptions (such as describing it as dragging or sliding along the skin). One subject described the sensation as \say{fun} and two described it as \say{nice}. 

Three participants described the sensation as a finger or hand moving along the skin. Two different participants described the sensation as a pencil running along their skin. Another three participants described the sensation like a tool or a toy being dragged on their forearm. Notably, six participants described the material as feeling like rubber. Additionally, six out of 16 participants said that the sensation felt like something rolling along their skin. Some subjects noted that the sensation felt lighter at the beginning of the contact, as the tactor moved into contact with the skin, and then felt lighter at the end, as the final tactor rolled off of the skin. Two of these six participants described the sensation as a large gear or wheel rolling along their skin. The following quotations represent some of the common themes and comments from participants in the study:
\begin{itemize}
    \item \say{It was a nice soft touch.}
    \item \say{It felt like around two fingertips width, and it was dragging from the top of my forearm to the bottom and then lifting off slowly.}
    \item \say{There wasn't enough friction such that my skin was being caught.}
    \item \say{It feels pretty consistent in force. So it feels pretty continuous. Like something is just sliding across my arm.}
\end{itemize}
Lastly, six participants used language that indicated they felt the sensation was \say{bumpy}. They described the sensation as \say{like a line of little dots but moving in sort of a wave} and \say{it has little nubbins but it felt like little nubbins that turned.} 

}
\subsection{\review{Discussion}}
\label{paradigm discussion}
\review{
    The results indicate that the majority of participants believe that a single contact surface is moving along the skin in a continuous motion. This supports our hypothesis that we can successfully create an illusion of continuous movement using discrete tactors. It is important to note, however, that it was common for some users to describe the sensation as an object rolling along their skin while others described it as an object dragging or sliding along the skin. Work done on the fingertip by Provancher et al. shows that it is possible to create  a virtual object with different radii of curvatures using rolling sensations \cite{provancher2005perception}. Our results indicate that it may be possible that their results extend to the forearm so that one could display round virtual objects to the forearm. However, our results also raise questions about exactly what attribute of the signal causes participants to believe the device is rolling on the skin. Participant responses indicate that the rolling sensations may be created by the way the tactors make and break contact with the skin but future work needs to be done to determine the exact mechanism.
    
    Although some participants described the sensation as \say{bumpy}, it was clear, in most cases, from their language and contextual clues from their statements that they thought that the object in contact and moving along their skin was bumpy, but that the movement itself was continuous. This may indicate that what participants feel is continuous is subjective and that potentially the optimal control parameters may not be the same for all participants because of variation in tactile sensitivity across the subject pool.
    
    After the open response study (and an additional study that will be described in Section~\ref{sec:distance}) was completed, participants were allowed to see the device. 15 of 16 participants wanted to see the device and of those 15, 14 were surprised to see that the device was composed of discrete elements because they had believed that there was only a single object that was in contact with their skin. This observation leads to questions about whether seeing the device and the mechanism creating the sensation biases user responses or effects their perception of continuity. 
}

\begin{figure}[t]
	\centering
	\includegraphics[width=0.8\columnwidth]{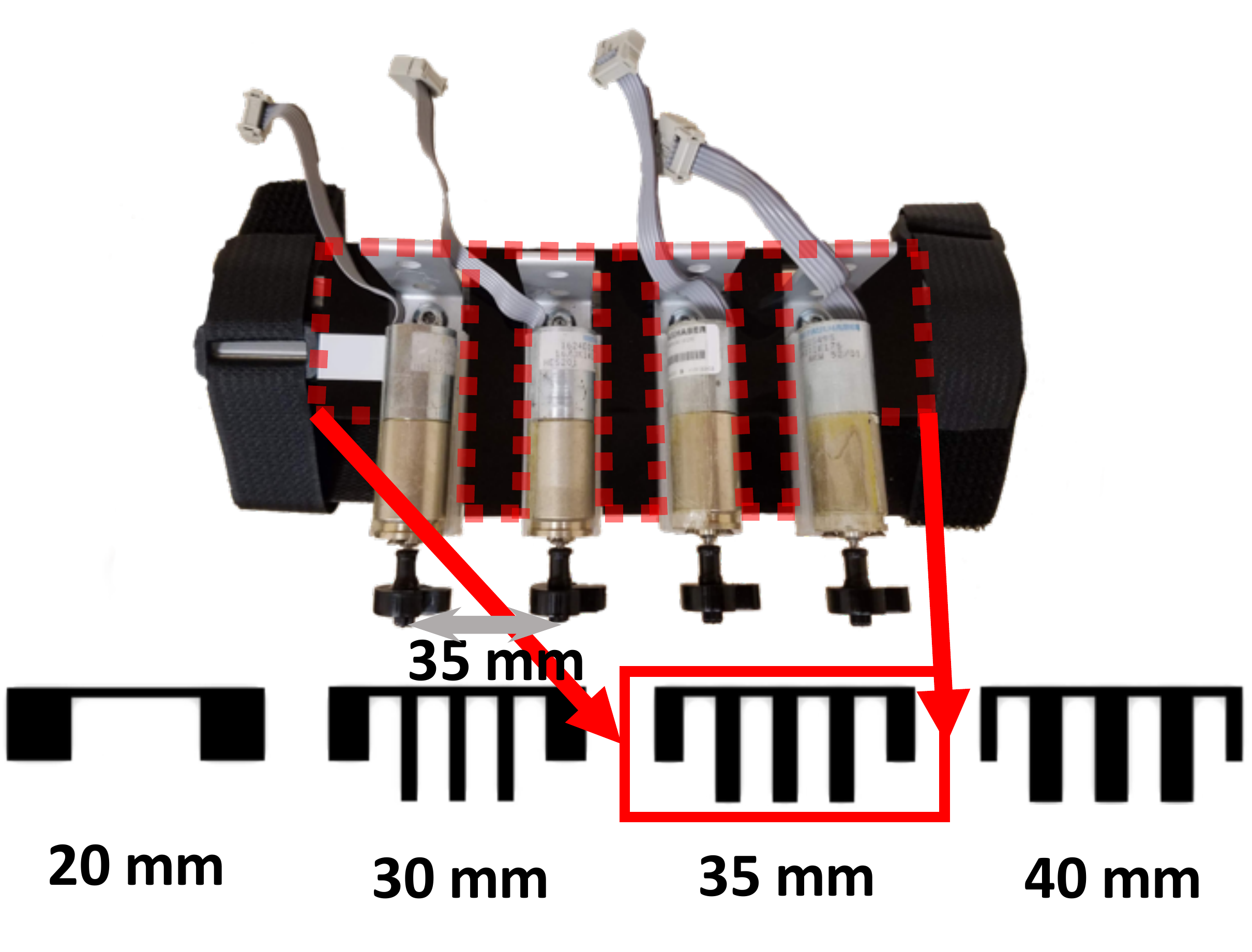}
	\vspace{-0.2in}
	\caption{The designs for the laser cut motor carriage spacers and an example showing how it is inserted to separate the motors at consistent and accurate distances.}\label{fig:ORP separators}
	\vspace{-0.1in}
\end{figure}

%\begin{table}
%\caption{Latin Square Design for Sequence of Contact Spacing}\label{table:latinsquares}
%\vspace{-0.5cm}
%\begin{center}
%\begin{tabular}[b!]{ |c|c|c|c|c| } 
% \hline
% & \bfseries Trial 1 & \bfseries Trial 2 & \bfseries Trial 3 & \bfseries Trial 4\\ 
% \hline
% \bfseries \review{Subjects 1 - 4} & 20~mm & 30~mm & 35~mm & 40~mm \\ 
% \hline
% \bfseries \review{Subjects 5 - 8} &30~mm & 20~mm & 40~mm & 35~mm\\ 
% \hline
% \bfseries \review{Subjects 9 - 12} &35~mm & 40~mm & 20~mm & 30~mm\\ 
% \hline
% \bfseries \review{Subjects 13 - 16} &40~mm & 35~mm & 30~mm & 20~mm\\ 
% \hline
%\end{tabular}
%\end{center}
%%\vspace{0.08in}
%\vspace{-0.2in}
%\end{table}

%\section{Open Response Contact Spacing Study}
\section{Contact Spacing \review{User} Study}
\label{sec:distance}

\begin{figure*}[t]
	\centering
	\includegraphics[width=2\columnwidth]{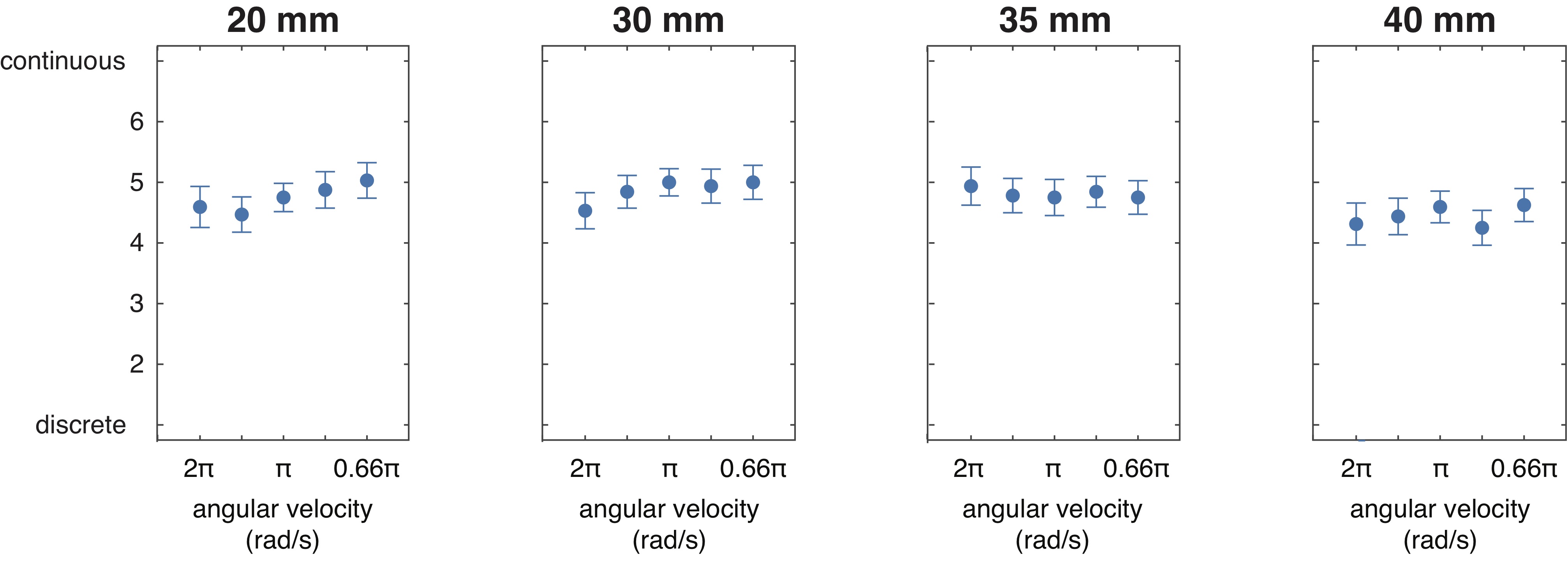}
	\vspace{-0.1in}
	\caption{\review{Average continuity ratings of all participants with standard error bars.}} %Linear regression was done on the average ratings ($C$ is the continuity, $T$ is time).}}
	\label{fig:contGraphDist}
	\vspace{-0.1in}
\end{figure*}
\begin{figure*}[t]
	\centering
	\includegraphics[width=2\columnwidth]{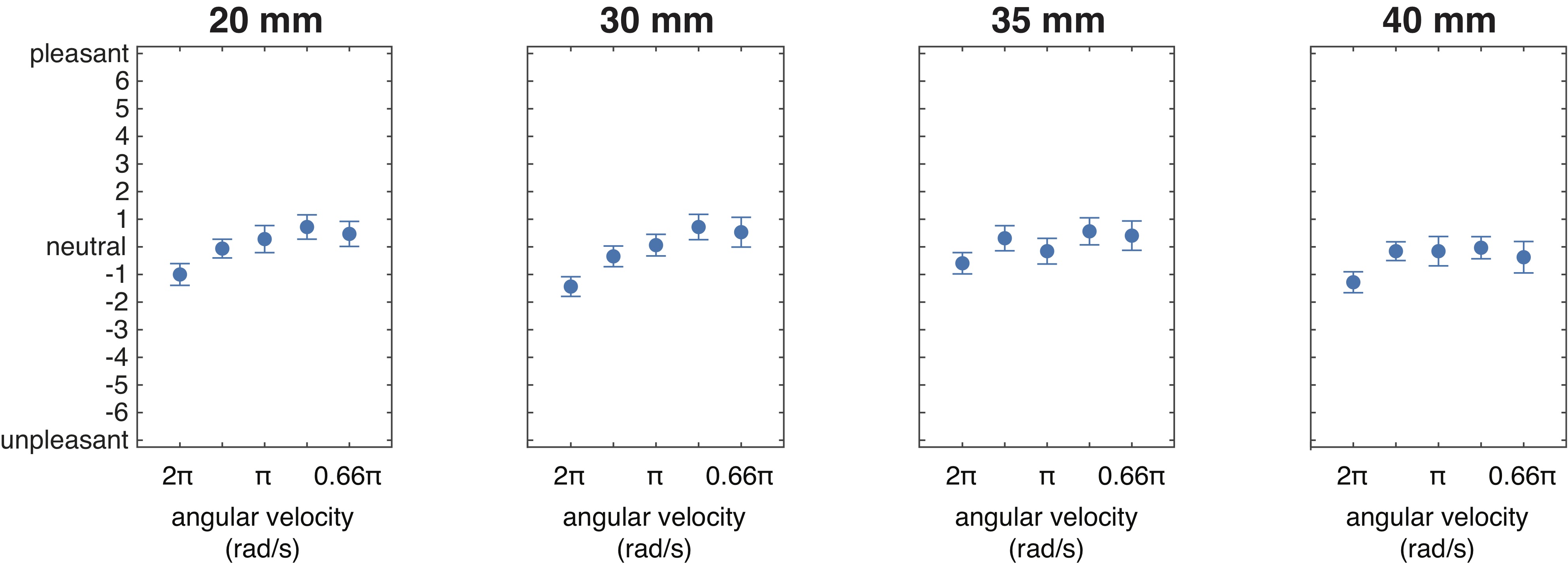}
	\vspace{-0.1in}
	\caption{\review{Average pleasantness ratings of all participants with standard error bars.}} %Linear regression was done on the average ratings ($P$ is the pleasantness, $T$ is time).}}
	\label{fig:pleasGraphDist}
	\vspace{-0.1in}
\end{figure*}

To investigate the effect that spacing between contact points has on creating a continuous, pleasant sensation, we conducted \review{a user study with the same 16 subjects described in Section~\ref{sec:paradigm}. All subjects completed the open response study, but otherwise did not have any previous experience with the device.}
%an open response study with 4 subjects (all right-handed; 2 male, 2 female; aged 21-32). All 4 subjects were familiar with haptic devices, but none of the subjects had any previous experience with this haptic device or participated in the previous study.
The protocol was approved by the Stanford University Institutional Review Board, and all subjects gave informed consent.

\subsection{Methods}
\label{distance methods}
Subjects sat at a table and placed their wrist and elbow in the position shown in Fig.~\ref{fig:studysetup}, which matches the positioning from the user study for the tactors contacting the volar forearm \review{described in detail and used in Section~\ref{sec:methods} and also Section~\ref{sec:paradigm}}. The participants were blindfolded and wore noise-canceling headphones playing white noise so that they could not see or hear the motors and tactors, nor see what changes were being made to the system during the study. \review{Participants did not remove the blindfold or the noise-canceling headphones until they had completed the study and therefore had their senses impaired both during the trials and the intervals between trials.}

%The user study described in Section~\ref{sec:study} showed that there was no significant difference in continuity between the dorsal and volar forearm, but that the sensations felt on the volar forearm were significantly more pleasant that those felt on the dorsal forearm. Additionally, users reported feeling more comfortable using their volar forearm to interact with the device than with their dorsal forearm. Thus, we chose the underside of the forearm as the contact location for the open response study.

Spacing between contact points in this investigation is defined as the distance between the centers of the shafts of the motors. In the previous user study \review{and open response study}, the contact points of the tactors were equally spaced every 20~mm \review{($D$ = 20~mm).} %and thus, 20~mm spacing was used as the baseline for the open response study. 
Since we were interested in the effect that increased distance between contact points would have on the sensation, we varied the spacing of the contact points between \review{$D$ = 20~mm, 30~mm, 35~mm, and 40~mm}. As discussed in Section~\ref{paradigm methods}, $N$ = 4 such that the device could still fit within the workspace.
%While the original design of the haptic device used in the previous user study consisted of an array of five Falhauber motors \review{($N$ = 5)}, we adapted the haptic device to be an array of four motors \review{($N$ = 4)} such that the device could still fit within the workspace (volar forearm) when increasing the spacing.
%In using four motors instead of five we were also able to gain some insight pertaining to the effect that the number of contact points has on the sensation. 
In order to be able to quickly and accurately change the spacing between contact points, we designed and laser cut motor spacers out of 1/4-inch acrylic (Fig.~\ref{fig:ORP separators}) that are inserted between and around the motor carriages.% during each trial.

\review{In addition to varying the distance between contact points, we also varied the \secondreview{angular velocity (2$\pi$, 1.33$\pi$, $\pi$, 0.8$\pi$, and 0.66$\pi$~rad/s). }%duration of tactor rotation (1.0, 1.5, 2.0, 2.5, and 3.0 seconds).
The delay was held constant at 10\% for all trials because this delay corresponds to the most continuous and pleasant sensations. These parameters resulted in 20 unique actuation conditions, each of which was displayed twice for a total of 40 trials. The participants completed the 40 trials in sets of 10 trials. Each set of 10 trials was conducted at a specific distance value. The order of duration of tactor rotation was randomized within each set of 10 trials. The sequence order for the contact spacing followed a Latin Square Design. % the Latin Square Design shown in Table~\ref{table:latinsquares}.
After 10 trials at a set distance, participants were given a 2 minute break during which the participants' forearms were taken out of the device and the spacing was changed. Participants were unaware that the break corresponded to changing the distance parameter.}

\review{After each trial, participants rated the sensation on its perceived continuity and pleasantness. Participants rated continuity using a 7-point Likert scale (1=Discrete and 7=Continuous). Similarly, they rated pleasantness on a Likert scale ranging from -7 to +7 (-7=Very Unpleasant, 0=Neutral, +7=Very Pleasant). After completing all 40 trials, participants completed a post-study survey which asked participants to rate using a 7-point Likert scale how difficult it was to distinguish sensations between trials, asked participants to provide information regarding how to they differentiated between trials, and were also given space to provide any additional comments. On average, the participants completed the study in 30 minutes.}

\subsection{Results}
\label{distance results}
\review{
Figure~\ref{fig:contGraphDist} shows the average continuity rating across all participants, separated by distance and \secondreview{angular velocity. }%duration of rotation. We fit a linear regression to the average continuity ratings for each delay. 
We ran a two-way repeated measures ANOVA on the continuity ratings with distance and \secondreview{angular velocity }%duration of rotation
as factors. \secondreview{We found that both distance %delay
and angular velocity violated the assumption of sphericity so we used the lower-bound estimate of $\varepsilon$ = 0.33 to correct our calculations. }
This analysis showed that there was no significant difference in continuity ratings between the distances \secondreview{($F(1,204.6)=2.14$, $p=0.145$, ${\eta_p}^{2} = 0.010$) }%($F(3,620)=2.14$, $p=0.094$, ${\eta_p}^{2} = 0.010$)
or between the \secondreview{angular velocity }%duration of rotation
\secondreview{($F(1.32,204.6)=0.53$, $p=0.516$, ${\eta_p}^{2} = 0.003$). }
%($F(4,620)=0.53$, $p=0.713$, ${\eta_p}^{2} = 0.003$).
The interaction between distance and \secondreview{angular velocity }%duration of rotation
was also not significant \secondreview{($F(3.96,204.6)=0.32$, $p=0.863$, ${\eta_p}^{2} = 0.006$)}.%($F(12,620)=0.32$, $p=0.986$, ${\eta_p}^{2} = 0.006$).

Figure~\ref{fig:pleasGraphDist} shows the average pleasantness rating across all participants, separated by distance and \secondreview{angular velocity. }%duration of rotation.We fit a quadratic to the average pleasantness ratings for each delay.
We ran a two-way repeated measures ANOVA on the continuity ratings with distance and \secondreview{angular velocity }%duration of rotation
as factors. \secondreview{Again, we found that both distance %delay
and angular velocity violated the assumption of sphericity so we used the lower-bound estimate of $\varepsilon$ = 0.33 to correct our calculations. }
This analysis showed no significant difference in \finalreview{pleasantness }%continuity
ratings between the distances \secondreview{($F(1,204.6)=1.38$, $p=0.242$, ${\eta_p}^{2} = 0.007$) }%($F(3,620)=1.38$, $p=0.249$, ${\eta_p}^{2} = 0.007$)
but did show a significant difference between the \secondreview{angular velocity }%duration of rotation
\secondreview{($F(1.32,204.6)=7.32$, $p=004$, ${\eta_p}^{2} = 0.045$). }
%($F(4,620)=7.32$, $p=9.17\times10^{-6}$, ${\eta_p}^{2} = 0.045$).
The interaction between distance %arm location
and angular velocity %delay value
was not significant \secondreview{($F(3.96,204.6)=0.37$, $p=0.828$, ${\eta_p}^{2} = 0.007$). }%($F(12,620)=0.37$, $p=0.973$, ${\eta_p}^{2} = 0.007$). 
After running a post-hoc pairwise comparison test with a Bonferroni correction, we found that the pleasantness ratings for \secondreview{an angular velocity of 2$\pi$~rad/s was }%a rotation duration of 1.0~s were
significantly less than the pleasantness ratings of all of the other \secondreview{angular velocities }%rotation durations
($p < 0.01$). However, pleasantness for the other four \secondreview{angular velocities (1.33$\pi$, $\pi$, 0.8$\pi$, and 0.66$\pi$~rad/s) }%rotation durations (1.5, 2.0, 2.5, and 3.0~s)
was not statistically significantly different from each other ($p>0.05 $).}

%All subjects described the sensations as moving up the arm. One explicitly stated that it felt ``continuous", while the others implied that it felt continuous via their descriptions (such as calling it a smooth movement or sliding along the skin). Similarly, all subjects described the sensations as feeling pleasant. Again, one subject used the word ``pleasant" to describe the sensations, while the other subjects implied it via their descriptions. One subject stated that the sensation felt like human touch.

%In general, subjects were unable to vocalize differences between first and second actuations in a trial or any differences between the sensations they felt across all trials. None of the subjects were able to identify that we were changing the distances between the contact points. Subjects felt as though maybe there were differences between the sensations but couldn't quite identify what felt different. Some subjects thought that maybe there was a difference in the speed of the movement and others thought that the width (not length) of the sensation varied.
%\vspace{-0.75cm}
\subsection{Discussion}
\label{distance discussion}
The results of this study further demonstrate that we can create a pleasant, continuous linear sensation using discrete lateral skin-slip. These results also indicate that there is a negligible change in the sensation by increasing the distance between contact points on the skin.

Although we hypothesized that the illusion of a continuous lateral motion would disappear the further apart the contact points separated, subjects still felt that the sensation was continuous and pleasant even at double the distance of the original study and with one fewer contact point. A previous investigation showed that the sensitivity to two-point light touch stimuli on the forearm averaged between 30.7-35.9~mm~\cite{Nolan1982twoptdiscrimination}, so we believed that subjects should begin to notice a difference using spacing of 30, 35, and 40~mm. However, we hypothesize that because we are creating a more complex sensation than what is done in a two-point discrimination test, the absolute threshold must be larger than the distance values that we tested. Future work must be done in order to determine that threshold specifically.

\review{In the post-study surveys, all subjects mentioned using the speed of the sensation as their main method for distinguishing between trials and providing their rating for continuity and pleasantness. The apparent speed of the sensation varied from trial to trial depending on the combination of distance and duration of rotation. As shown in Table~\ref{table:orp}, the apparent speeds ranged from 4.8 - 26.7~cm/s. Increasing the spacing between contact points increases the apparent speed when delay, \secondreview{angular velocity, }%duration of pulse rotation,
and $N$ are held constant. A hypothesis for why the illusion of a continuous sensation was not broken as we increased the distance between contact points is that our apparent speeds were close to the plateau range, as shown in Fig.~\ref{fig:contApparentVelocity} and discussed more in detail in Section~\ref{sec:discussion}. We have yet to investigate whether the change in $D$ is perceptible with smaller apparent speeds or for other durations of rotation. Further experiments would need to be conducted to see if the trends found in this study are consistent invariant to other parameter changes.}

\review{The continuity values for this study were noticeably higher than for the initial user study. We hypothesize that this is because participants were blindfolded for the entirety of the study and never saw the device. Since we cannot run statistical tests on the data to determine if viewing the device is statistically significant because of the different subject pools, we believe that conducting a study in the future to specifically look at this effect will provide interesting contributions to the research field.}
\secondreview{We also believe that values for both studies may be slightly low due to participants' hesitation to rate values too far on one end of the scale. This effect was previously seen in a contact realism study in which participants did not rate tapping on physical wood as perfectly realistic compared to the sensation of tapping on wood~\cite{Kuchenbecker2006Realism}.}

Similarly, future work must also be done to determine the minimum number of contact points that are necessary in order to still feel the illusion of a continuous and pleasant linear sensation. The voice coil haptic device used in our previous work on normal skin indentation~\cite{VCstudy} had six contact points to create a continuous stroking sensation. Our investigation has shown that fewer contact points spaced further apart can still create a pleasant, continuous linear sensation\review{, as our initial user study used 5 contact points ($N$ = 5) and then our user study focusing on distance used 4 contact points ($N$ = 4).} This information is important for the design of wearable haptics because fewer actuators can be used which can reduce both the size of the device and the amount of power needed to actuate the device.

\begin{table}
\caption{\review{Computed Apparent Speeds of Contact Point ($N$ = 4 and $d=10$\%)}}\label{table:orp}
\vspace{-0.5cm}
\begin{center}
\includegraphics[width=\columnwidth]{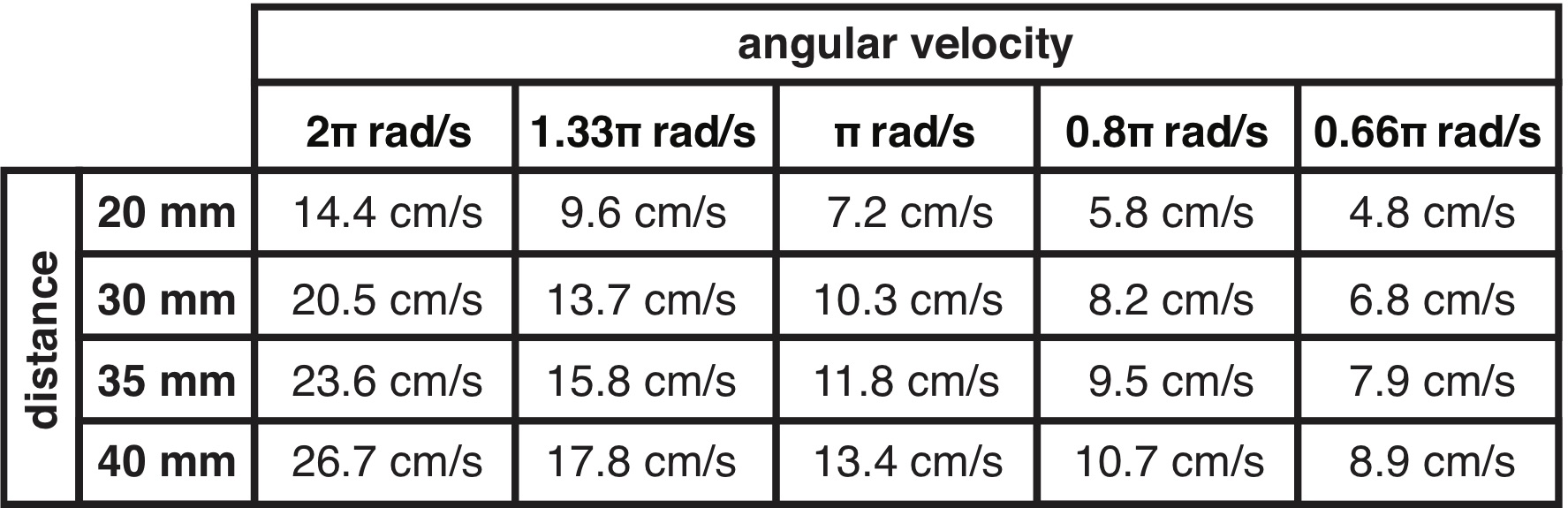}
	\vspace{-0.35in}
    \end{center}
\end{table}

\section{Conclusions and Future Work}
\label{sec:conclusion}
In this paper we presented the design and evaluation of a device that creates a pleasant, continuous linear sensation on the forearm using discrete lateral skin-slip. \review{The device could primarily be used to convey social touch cues, such as comfort and affection, in which stroking motions are used. Because the device can convey the movement of a contact point on the skin, it could also be used in virtual reality applications to convey touch interactions between individuals and virtual objects in cases where the travel of a contact point over the skin is relevant or for navigation tasks}. Users place either their volar or dorsal forearm into the world-grounded haptic device comprised of a linear array of motors which rotate to provide discrete lateral skin-slip. We first conducted a human-subject study to investigate the effect that the delay between the onset of actuation of the motors and the \secondreview{angular velocity }%duration of rotation
has on the sensation. We found that to optimize for a continuous sensation, one should command the motors with minimal delay and \secondreview{slow angular velocity. } %maximal duration of rotation.
To optimize for pleasantness, one should command the motors with medial delay values and a medial \secondreview{angular velocity. }%duration of rotation.
We then conducted \review{an open response study to determine if subjects would identify and describe the sensation as continuous and pleasant without being asked to rate the sensation on corresponding Likert scales. We then conducted a follow-up user study to }%a follow-up study in the form of an open response study to
investigate the effect that spacing between contact points has on the illusion of a continuous stroking sensation. Subjects were unable to discern any difference in the sensation, even when the spacing between contact points doubled. In future work, we will work to determine the threshold for spacing between contact points and also the effect that skin contact area has on the sensation.

It is possible that the values for continuity are artificially low \review{in the first study} because participants are able to see that the device was composed of separate actuators. \review{In our second user study comparing distance between contact points, the users were blindfolded such that they could not see the device and we received higher continuity ratings.} In future work, we will explore the extent of the illusion by performing a study with blinded participants comparing a device that actually performs continuous skin slip along the skin\review{, such as a robotic finger dragging along the skin,} and the discrete device presented here. 

The results from this paper show that it is possible to effectively create an illusory sensation of continuous lateral motion using discrete lateral skin-slip. This actuation method could be used to relay simple messages, including those pertaining to social touch and navigation. The results obtained during our investigations will help to inform the design of future wearable haptic devices and could help to reduce the overall size and mechanical complexity. Although our device was world-grounded and used large motors, the principles of discrete lateral skin-slip could be applied to a body-grounded wearable device with smaller actuators. The main requirements of these actuators would be to provide at least 0.3~mN of normal force and 1.5~mm of normal indentation that transitions to skin-slip with an apparent \review{speed} %velocity
$\approx$5-10~cm/s, as calculated using at least 4 contact points equally spaced no more than 40~mm apart.

Large arrays of actuators can be used to deliver haptic sensations on locations all over the body. However, rendering and control methods for actuating these arrays is still unclear. This paper introduces a rendering algorithm to create convincing stroking sensations using skin-slip that could be used for portable and wearable devices. \review{The use of haptic illusions, such as this one, will allow haptic designers to use smaller, lightweight actuators when creating wearable devices.} Additionally, the presented results provide a framework for designing a wearable device to perform skin-slip on the forearm. Further, the work and methods from this investigation can be extended to understand this form of feedback on other locations on the body.

\section*{Acknowledgements}
The authors would like to thank Ali Israr, Frances Lau, Keith Klumb, and Freddy Abnousi, for their collaboration on methods to convey information through lateral skin stimulation.

\bibliographystyle{IEEEtran}
\bibliography{Rollers}

\vfill
%\vskip -2\baselineskip plus -1fil

\begin{IEEEbiography}[{\includegraphics[width=1in,height=1.25in,clip,keepaspectratio]{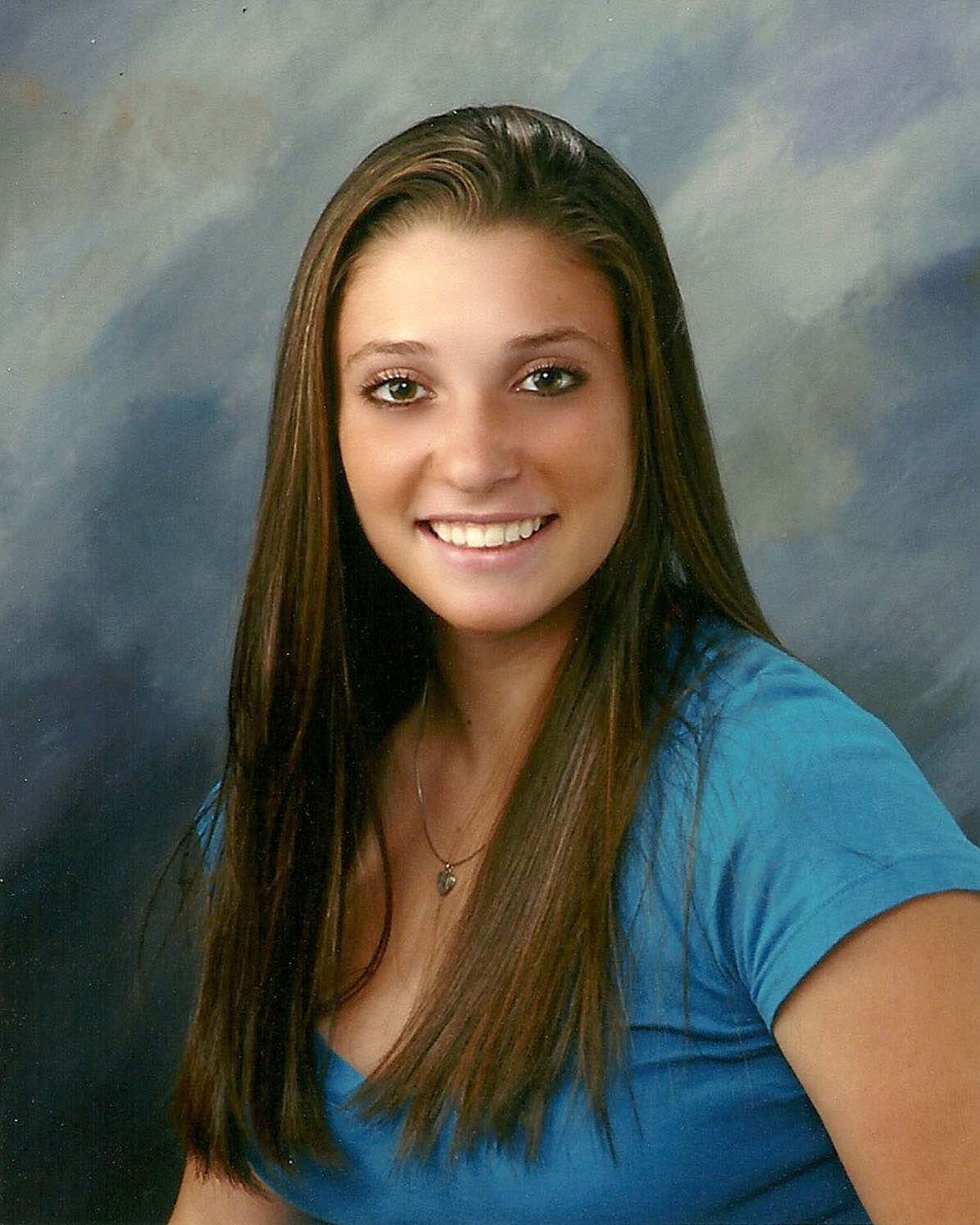}}]
{Cara M. Nunez} is a PhD candidate at Stanford University in the Collaborative Haptics and Robotics in Medicine Lab. She received a M.S. in mechanical engineering from Stanford University in 2018, and a B.S. in biomedical engineering and a B.A. in Spanish as a part of the International Engineering Program from the University of Rhode Island in 2016. Her research interests include haptics and robotics, with a specific focus on haptic perception, cutaneous force feedback techniques, and wearable devices, for medical applications, human-robot interaction, virtual reality, and STEM education. She is a National Science Foundation Graduate Research Fellow and a student member of the IEEE.
\end{IEEEbiography}

%\vfill
%\vskip 0pt plus -1fil
%\vskip -2\baselineskip plus -1fil
\vskip -2\baselineskip plus -1fil

\begin{IEEEbiography}[{\includegraphics[width=1in,height=1.25in,clip,keepaspectratio]{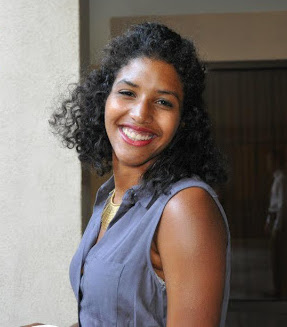}}]
{Sophia R. Williams} is a PhD Candidate in the Collaborative Haptics and Robotics in Medicine (CHARM) Lab at Stanford University. Sophia received her B.S. from Harvey Mudd College in 2015. Her research interests include wearable haptic interfaces, psychophysics, optimization and computer science for design applications, medical robots, and human-robot interaction. Sophia is a National Science Foundation Graduate Research Fellow and recipient of the Watson Fellowship.
\end{IEEEbiography}

%\vfill
%\vskip 0pt plus -1fil
\vskip -2\baselineskip plus -1fil

\begin{IEEEbiography}[{\includegraphics[width=1in,height=1.25in,clip,keepaspectratio]{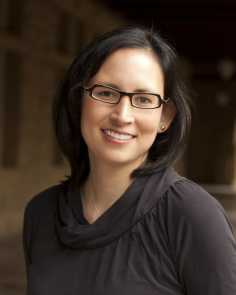}}]
{Allison M. Okamura} received the B.S. degree from the University of California, Berkeley, in 1994, and the M.S. and Ph.D. degrees from Stanford University, Stanford, CA, in 1996 and 2000, respectively, all in mechanical engineering. She is currently a Professor of mechanical engineering at Stanford University, Stanford, CA. Her research interests include haptics, teleoperation, medical robotics, virtual environments and simulation, neuromechanics and rehabilitation, prosthetics, and engineering education.
\end{IEEEbiography}

%\vfill
%\vskip 0pt plus -1fil
\vskip -2\baselineskip plus -1fil

\begin{IEEEbiography}
[{\includegraphics[width=1in,height=1.25in,clip,keepaspectratio]{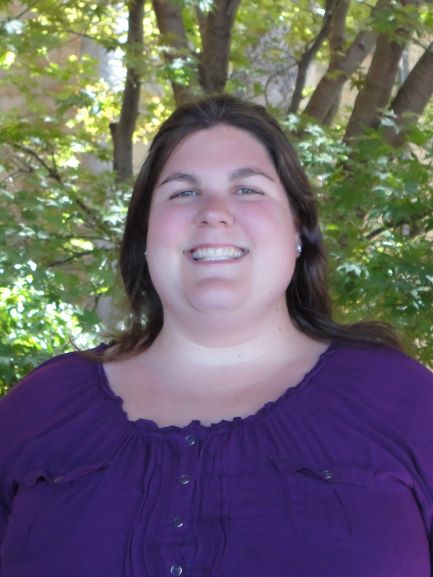}}]{Heather Culbertson}
is an assistant professor of Computer Science and Aerospace and Mechanical Engineering at the University of Southern California. She received the Ph.D. degree in the department of mechanical engineering and applied mechanics (MEAM) at the University of Pennsylvania in 2015. She completed the M.S. degree in MEAM at the University of Pennsylvania in May of 2013, and she earned the B.S. degree in mechanical engineering at the University of Nevada, Reno in 2010. Prior to joining USC she was a research scientist at Stanford University. Her research focuses on the design and control of haptic devices and rendering systems, human-robot interaction, and virtual reality.
\end{IEEEbiography}

%\vfill

\end{document}